\documentclass[journal]{IEEEtran}
\ifCLASSINFOpdf
\else
\fi

\usepackage{amsmath}
\usepackage{epsfig}
\usepackage{graphicx}
\usepackage{multirow}
\usepackage{subcaption}
\usepackage{mwe}
\usepackage{makecell}
\usepackage{url}

\newcommand{\comment}[1]{}

\usepackage{color}

\begin{document}
\title{Multi-Sensor Data Fusion for Cloud Removal in Global and All-Season Sentinel-2 Imagery}

\author{Patrick~Ebel,~\IEEEmembership{Student Member,~IEEE,}
        Andrea~Meraner,~%~\IEEEmembership{TODO,~OSA,}
        Michael~Schmitt,~\IEEEmembership{Senior Member,~IEEE,}
        and~Xiao~Xiang~Zhu,~\IEEEmembership{Senior Member,~IEEE}% <-this % stops a space
%\thanks{M. Shell is with the Department
%of Electrical and Computer Engineering, Georgia Institute of Technology, Atlanta,
%\thanks{Manuscript received April 19, 2005; revised September 17, 2014.}

\thanks{This work was partially supported by the Federal Ministry for Economic Affairs and Energy of Germany in the project “AI4Sentinels– Deep Learning for the Enrichment of Sentinel Satellite Imagery” (FKZ50EE1910). The work of X. Zhu is jointly supported by the European Research Council (ERC) under the European Union's Horizon 2020 research and innovation programme (grant agreement No. [ERC-2016-StG-714087], Acronym: \textit{So2Sat}), by the Helmholtz Association
through the Framework of Helmholtz Artificial Intelligence Cooperation Unit (HAICU) - Local Unit ``Munich Unit @Aeronautics, Space and Transport (MASTr)'' and Helmholtz Excellent Professorship ``Data Science in Earth Observation - Big Data Fusion for Urban Research'', and by the German Federal Ministry of Education and Research (BMBF) in the framework of the international future AI lab ``AI4EO -- Artificial Intelligence for Earth Observation: Reasoning, Uncertainties, Ethics and Beyond''. 
(\textit{Corresponding Author: Xiao Xiang Zhu})} 

% Michael's formulation:
\thanks{P. Ebel is with Signal Processing in Earth Observation, Technical
University of Munich, 80333 Munich, Germany. (e-mail: patrick.ebel@tum.de).}% <-this % stops a space
\thanks{A. Meraner is with EUMETSAT European Organisation for the Exploitation of Meteorological Satellites, 64295 Darmstadt. This work is done, when he was with Signal Processing in Earth Observation, Technical University of Munich (e-mail: andrea.meraner@eumetsat.int).}% <-this % stops a space
\thanks{M. Schmitt is with the Department of Geoinformatics, Munich University of Applied Sciences. This work is done, when he was with Signal Processing in Earth Observation, Technical University of Munich (e-mail: michael.schmitt@hm.edu).}% <-this % stops a space
\thanks{X. X. Zhu is with the Remote Sensing Technology Institute, German Aerospace Center, 82234 Wessling, Germany, and also with the Signal Processing in Earth Observation, Technical University of Munich, 80333 Munich, Germany. (e-mail: xiaoxiang.zhu@dlr.de).}% <-this % stops a space

% Xiaoxiang's formulation:
%\thanks{P. Ebel, A. Meraner and M. Schmitt are with Signal Processing in Earth Observation, Technical
%University of Munich, 80333 Munich, Germany. (e-mails: patrick.ebel@tum.de, andrea.meraner@eumetsat.int, michael.schmitt@hm.edu)}% <-this % stops a space
%\thanks{X. X. Zhu is with the Remote Sensing Technology Institute, German Aerospace Center, 82234 Wessling, Germany, and also with the Signal Processing in Earth %Observation, Technical University of Munich, 80333 Munich, Germany. (e-mail: xiaoxiang.zhu@dlr.de)}% <-this % stops a space
%\thanks{Present address of A. Meraner: EUMETSAT European Organisation for the Exploitation of Meteorological Satellites, 64295 Darmstadt.}
%\thanks{Present address of M. Schmitt: Department of Geoinformatics, Munich University of Applied Sciences, 80333 Munich.}

}

\markboth{IEEE TRANSACTIONS ON GEOSCIENCE AND REMOTE SENSING, VOL. TODO, NO. TODO, TODO TODO}%
{Shell \MakeLowercase{\textit{et al.}}: Bare Demo of IEEEtran.cls for Journals}



\maketitle

\begin{abstract}

    This work has been accepted by IEEE TGRS for publication. The majority of optical observations acquired via spaceborne earth imagery are affected by clouds. While there is numerous prior work on reconstructing cloud-covered information, previous studies are oftentimes confined to narrowly-defined regions of interest
    , raising the question of whether an approach can generalize to a diverse set of observations acquired at variable cloud coverage or in different regions and seasons. We target the challenge of generalization by curating a large novel data set for training new cloud removal approaches and evaluate on two recently proposed performance metrics of image quality and diversity. Our data set is the first publically available to contain a global sample of co-registered radar and optical observations, cloudy as well as cloud-free. Based on the observation that cloud coverage varies widely between clear skies and absolute coverage, we propose a novel model that can deal with either extremes and evaluate its performance on our proposed data set. 
    Finally, we demonstrate the superiority of training models on real over synthetic data, underlining the need for a carefully curated data set of real observations.
    To facilitate future research, our data set is made available online\footnote{The SEN12MS-CR data set is publicly available online under \url{https://mediatum.ub.tum.de/1554803}.}.
\end{abstract}


% \fi
\IEEEpeerreviewmaketitle

\section{Introduction}
%\IEEEPARstart{T}{his} demo file is intended to serve as a ``starter file'' for IEEE journal papers produced under \LaTeX\ using IEEEtran.cls version 1.8a and later.
% I wish you the best of success.
%\hfill September 17, 2014

\IEEEPARstart{O}{n} average about $55$ \% of the earth's land surface is covered by clouds \cite{King_Platnick_Menzel_Ackerman_Hubanks_2013}, impacting the aim of missions like Copernicus to reliably provide noise-free observations at a high frequency, a prerequisite for applications relying on temporally seamless monitoring of our environment, such as change detection or monitoring \cite{singh1989review, jensen1996introductory, coppin2002digital, Woodcock_Loveland_Herold_Bauer_2019}. The need for cloud-free earth observations hence gave rise to a rapidly growing number of cloud removal methods \cite{Enomoto_Sakurada_Wang_Fukui_Matsuoka_Nakamura_Kawaguchi_2017, Grohnfeldt_Schmitt_Zhu_2018, Singh_Komodakis_2018,  Bermudez_Happ_Feitosa_Oliveira_2019, Rafique_Blanton_Jacobs, Sarukkai_Jain_Uzkent_Ermon_2019, Meraner}. While the aforementioned contributions share the common aim of dehazing and declouding optical imagery, the majority of methods are evaluated on narrowly-defined and geo-spatially distinct regions of interest. 
Not only is this specificity posing challenges for a conclusive comparison of methodology but, furthermore, may cloud-removal performance on a particular region of interest poorly indicate performances on other parts of the globe or at different seasons. Moreover, it would be desirable for a cloud removal method to be equally applicable to all regions on earth, at any season. This generalisability would allow for large-scale earth observation without the need for costly re-designing or re-training for each individual scene that a cloud removal method is meant to be applied to. 

\begin{figure}[h!tb]
  \centering
  \begin{subfigure}[b]{0.3\linewidth}
    \includegraphics[width=\linewidth]{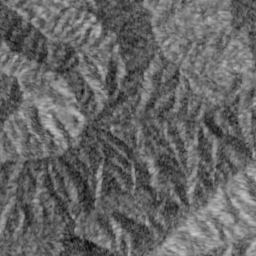}
  \end{subfigure}
  \begin{subfigure}[b]{0.3\linewidth}
    \includegraphics[width=\linewidth]{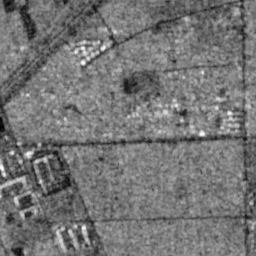}
  \end{subfigure}
    \begin{subfigure}[b]{0.3\linewidth}
    \includegraphics[width=\linewidth]{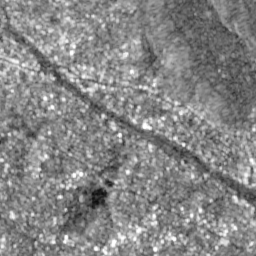}
  \end{subfigure}
  \label{fig:coffee}
  \begin{subfigure}[b]{0.3\linewidth}
    \includegraphics[width=\linewidth]{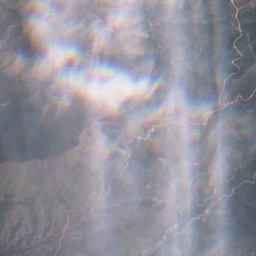}
  \end{subfigure}
  \begin{subfigure}[b]{0.3\linewidth}
    \includegraphics[width=\linewidth]{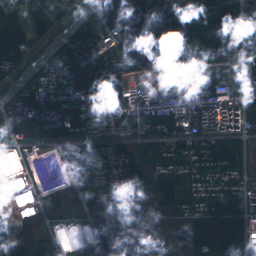}
  \end{subfigure}
    \begin{subfigure}[b]{0.3\linewidth}
    \includegraphics[width=\linewidth]{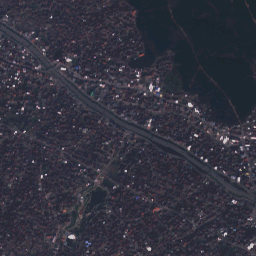}
  \end{subfigure}
  \begin{subfigure}[b]{0.3\linewidth}
    \includegraphics[width=\linewidth]{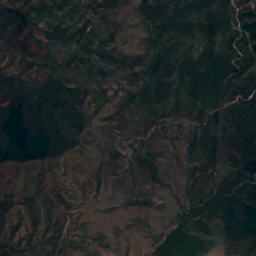}
  \end{subfigure}
  \begin{subfigure}[b]{0.3\linewidth}
    \includegraphics[width=\linewidth]{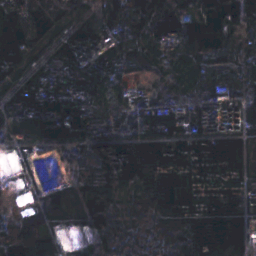}
  \end{subfigure}
    \begin{subfigure}[b]{0.3\linewidth}
    \includegraphics[width=\linewidth]{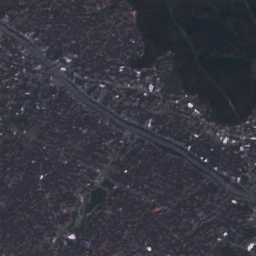}
  \end{subfigure}
  \begin{subfigure}[b]{0.3\linewidth}
    \includegraphics[width=\linewidth]{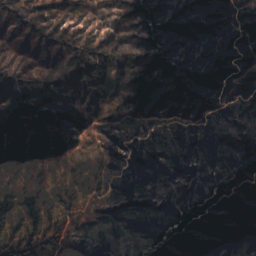}
  \end{subfigure}
  \begin{subfigure}[b]{0.3\linewidth}
    \includegraphics[width=\linewidth]{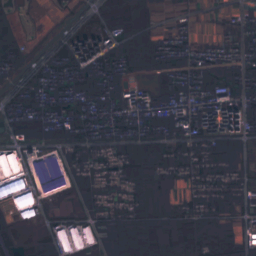}
  \end{subfigure}
    \begin{subfigure}[b]{0.3\linewidth}
    \includegraphics[width=\linewidth]{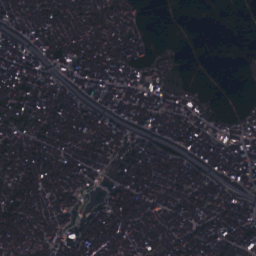}
  \end{subfigure}
  \caption{Exemplary raw data and de-clouded images. Rows: S1 data (in grayscale), S2 data (in RGB), predicted $\hat{S2}$ data, cloud-free (target) S2 data. Columns: three different samples. The outcomes show that our model learns to preserve optical data of cloudless areas while replacing cloudy regions by the translation from the SAR domain.
  }
  \label{fig:teaser}
\end{figure}

This concern is sustained by previous analysis demonstrating that landcover statistics differ across continents \cite{friedl2002global} and cloud-coverage is highly variable depending on meteorological seasonality \cite{King_Platnick_Menzel_Ackerman_Hubanks_2013}.
A major reason for these issues still remaining open nowadays is the current lack of available large-scale data sets for both training and testing of modern cloud removal approaches. In this work, we address this issue by curating and releasing a novel large-scale data set for cloud removal containing over 100{,}000 samples from over 100 regions of interests distributed over all continents and meteorological seasons of the globe. Specifically, we address the challenge of cloud removal in observations from Copernicus mission's Sentinel-2 (S2) satellites. While optical imagery is affected by bad weather conditions and lack of daylight, sensors based on synthetic aperture radar (SAR) as mounted on Sentinel-1 (S1) satellites are not \cite{bamler2000principles} and thus provide a valuable source of complementary information. Recent advances in cloud removal combine multi-modal data with deep neural networks recovering the affected areas \cite{Enomoto_Sakurada_Wang_Fukui_Matsuoka_Nakamura_Kawaguchi_2017, Grohnfeldt_Schmitt_Zhu_2018, Bermudez_Happ_Oliveira_Feitosa_2018, Meraner}. 
However, many networks are trained on synthetic data or on real data while making strong assumptions on the type and amount of cloud coverage. Moreover, the majority of methods do not explicitly model the amount of cloud coverage and treat each pixel similarly, thereby making unneeded changes to cloud-free areas. 

In this work we address the problem of cloud removal in optical data by means of SAR-optical data fusion. To redeem the current lack of sufficiently-sized and heterogeneous earth observation data for cloud removal we release a novel large-scale global data set of co-registered optical cloudy, cloud-free and SAR observations to train and test declouding methods on. Our data set consists of over a hundred thousand samples, allowing the training of large models for cloud removal and capturing a diverse range of observations from all continents and meteorological seasons. In addition, we propose a novel generative architecture that reaches competitive performance, as evidenced by two very recently proposed metrics of generated image goodness and diversity. Finally, we show that synthetic data utilized in previous studies is a poor substitute for real cloud coverage data, underpinning the needs for the novel data set proposed in our work. 

\subsection{Related Work}
The first deep neural architecture to reconstruct cloud-covered images combined near-infrared and red-green-blue (RGB) bandwith optical imagery by means of a conditional generative adversarial network (GAN) \cite{Enomoto_Sakurada_Wang_Fukui_Matsuoka_Nakamura_Kawaguchi_2017}, motivated by infrared bandwith being to a lesser extent impacted by cloud coverage. Subsequent studies replaced the infrared input with SAR observations \cite{Grohnfeldt_Schmitt_Zhu_2018, Bermudez_Happ_Oliveira_Feitosa_2018} due to SAR microwaves not being affected by clouds at all \cite{bamler2000principles}. While the early work of \cite{Enomoto_Sakurada_Wang_Fukui_Matsuoka_Nakamura_Kawaguchi_2017, Grohnfeldt_Schmitt_Zhu_2018} provide a proof-of-concept solely on synthetic data of simulated Perlin noise \cite{Perlin}, the networks of \cite{Bermudez_Happ_Oliveira_Feitosa_2018, Singh_Komodakis_2018} were first to demonstrate performances on real-world data, although focusing primarily on removal of filmy clouds. Comparable to these studies, we investigate the benefits of SAR-optical data fusion for cloud removal. Unlike the prior work, we address declouding on a carefully curated data set of real imagery sampled over all continents and meteorological seasons, relying neither on synthetic data nor making any strong assumptions about the type and percentage of cloud coverage. Building on the previous studies, the models of \cite{Singh_Komodakis_2018, Fuentes_Reyes_Auer_Merkle_Henry_Schmitt_2019} replace the conditional GAN by a cycle-consistent architecture \cite{Zhu_Park_Isola_Efros_2017}, relaxing the preceding models' requirements for pixel-wise corresponding training data pairs. While \cite{Singh_Komodakis_2018} relies solely on cloudy optical input data at inference time, only SAR observations are utilized in \cite{Fuentes_Reyes_Auer_Merkle_Henry_Schmitt_2019}. Similar to these two networks, the model we propose uses a cycle-consistent GAN architecture. We combine cloudy optical with SAR observations and extend on the previous models by incorporating a focus on local reconstruction of cloud-covered areas. This is in line with very recent work \cite{Gao_Yuan_Li_Zhang_Su_2020, Meraner} that proposed an auxiliary loss term to encourage the model reconstructing information of cloud-covered areas in particular. The network of \cite{Meraner} is noteworthy for two reasons: First, for departing from the previous generative architectures by using a residual network (ResNet) \cite{He_Zhang_Ren_Sun_2016} trained supervisedly on a globally sampled data set of paired data. Second, for adding a term to the local reconstruction loss that explicitly penalizes the model for modifying off-cloud pixels. Comparable to \cite{Meraner} our network explicitly models cloud coverage and minimizes changes to cloud-free areas. Unlike the model of \cite{Meraner} our architecture follows that of cycle-consistent GAN and has the advantage of not requiring pixel-wise correspondences between cloudy and non-cloudy optical training data, thereby also allowing for training or fine-tuning on data where such a requirement may not be met. Complementary to the SAR-optical data fusion approach to cloud removal, recent contributions proposed integrating information of repeated observations over time \cite{Rafique_Blanton_Jacobs, Sarukkai_Jain_Uzkent_Ermon_2019}. The work indicates promising results but trades temporal resolution for obtaining a single cloud free observation, whereas our approach predicts one cloud-free output per cloudy input image and thus allows for sequence-to-sequence translation. Moreover, current multi-temporal approaches make strong assumptions about the maximum permissible amount of cloud-coverage affecting individual images in the input time series, which is required to be no more than 25 or 50 \% of cloud coverage for the method of \cite{Rafique_Blanton_Jacobs} and 10-30 \% in the work of \cite{Sarukkai_Jain_Uzkent_Ermon_2019}. Our curated data set evidences that such strict requirements on the percentage of cloudiness may oftentimes not be met in practice. Consequently, our model makes no assumptions on the maximum amount of tolerable cloud coverage per observation and can gracefully deal with samples ranging from cloud-free to widely obscured skies, thanks to minimizing changes to cloud-free pixels and using SAR observations unaffected by clouds.

\begin{figure*}[h!tb]
    \centering
    \includegraphics[width=\linewidth]{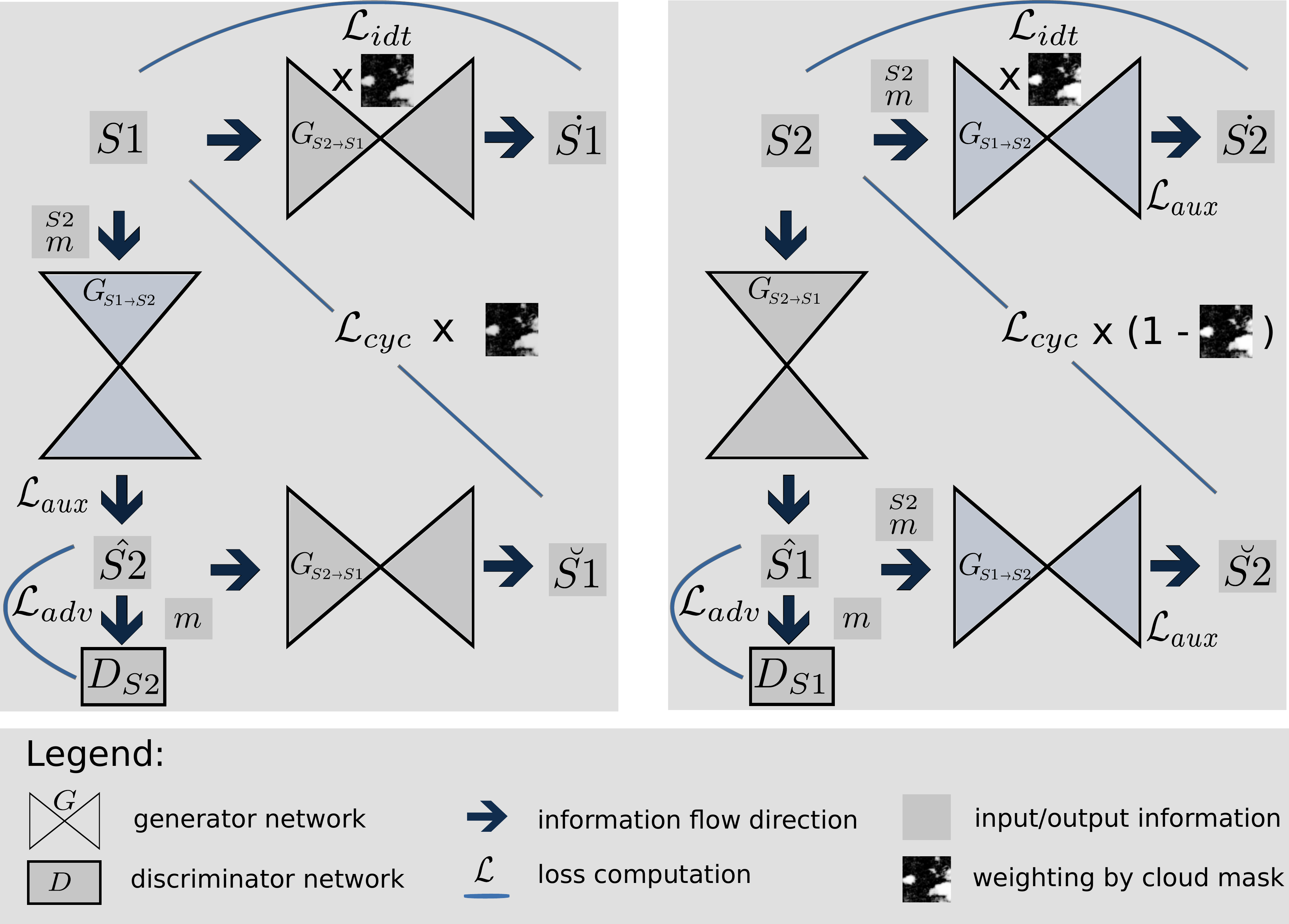}
    \caption{An overview of our model ensemble, based on cycle-consistent Generative Adversarial Networks \cite{Zhu_Park_Isola_Efros_2017}. The model consists of two generative networks $G_{S1 \rightarrow S2}$ and $G_{S2 \rightarrow S1}$ that translate images from the source domain of $S1$ to the target domain of $S2$ and vice versa. Distribution $\dot{S1}$ (or $\dot{S2}$) denotes the target when the generator performs a within-domain identity mapping, preserving the input image's sensor characteristics. For each domain there exists an associated discriminator network, denoted as $D_ {S1}$ and $D_{S2}$ respectively, classifying whether a given image is a sample from the domain's true distribution S1 (or S2) or from the synthesized distribution $\hat{S1}$ (or $\hat{S2}$). The network architectures are as in \cite{Zhu_Park_Isola_Efros_2017}---except for the generator $G_{S1 \rightarrow S2}$, which is modified as detailed in the main text and in Fig. \ref{fig:generator}. The losses $\mathcal{L}_{adv}, \mathcal{L}_{cyc}, \mathcal{L}_{idt}$ and $\mathcal{L}_{aux}$ are defined in section \ref{sub:losses} of the main text.} 
    \label{fig:architecture}
\end{figure*}

\section{Methods}\label{sec:methods}
We propose a novel model to recover cloud-occluded information in optical imagery. Our network explicitly processes a continuous-valued mask of cloud coverage computed on the fly as described in section \ref{sub:cloudmap} to preserve cloud-free pixels while making data-driven adjustments to cloudy areas.
The continues-valued assignment of each pixel in the processed cloud mask can be interpreted as the likelihood of the pixel being cloud-covered, according to the cloud detector algorithm of \cite{Zupanc}. Our model explicitly processing cloud coverage information is in contrast to previous generative architectures that are agnostic to cloud-coverage \cite{Enomoto_Sakurada_Wang_Fukui_Matsuoka_Nakamura_Kawaguchi_2017, Singh_Komodakis_2018} and networks that only utilize binary cloud mask information \cite{Meraner} as opposed to more fine-grained continuous-valued masks proposed in this work. A cycle-consistent generative architecture detailed in section \ref{sub:architecture} allows for training without the need for co-registered cloudy and non-cloudy observations of strict pixelwise one-to-one correspondences, as compared to earlier approaches that required strict pixel-wise alignments \cite{Bermudez_Happ_Oliveira_Feitosa_2018, Grohnfeldt_Schmitt_Zhu_2018}. We adapt the architecture to integrate SAR with optical observations and propose a new auxiliary cloud map regression loss that enforces sparse reconstructions to minimize modification on cloud-free areas as described in section \ref{sub:losses}. 

\subsection{Cloud detection \& mask computation} \label{sub:cloudmap}
To evaluate the cloud coverage statistics of our collected data set and model cloud coverage explicitly while reconstructing cloud-covered information, we compute cloud probability masks $m$. The masks $m$ are computed online for each cloudy optical image and contain continuous pixel values within $[0,1]$, indicating for a given pixel its probability of being cloud-covered. We compute $m$ via the classifier s2cloudless of \cite{Zupanc}, which demonstrated cloud detection accuracies on par with the multi-temporal classifier MAJA \cite{lonjou2016maccs}, running on single-shot observations. While s2cloudless originally applies classification to compute a sparsified binary cloud mask, we wish to obtain a continuous-valued cloud map. We therefore take the intermediate continuous-valued representation of the pipeline of \cite{Zupanc}, then apply a high-pass filter to only keep values above $0.5$ intensity and finally convolve with a Gaussian kernel of width $\sigma=2$ to get a smoothed cloud map with pixel values in [0, 1]. We note that $m$ may alternatively be computed by a dedicated deep neural network \cite{jeppesen2019cloud}, but our solution is lightweight and thus perfect to support methods running on very large data sets, at almost no additional computational cost in either memory or run time. Exemplary samples of cloud probability masks are presented in appendix \ref{appendix:a}.

\begin{figure*}[h!tb]
    \centering
    \includegraphics[width=1\linewidth]{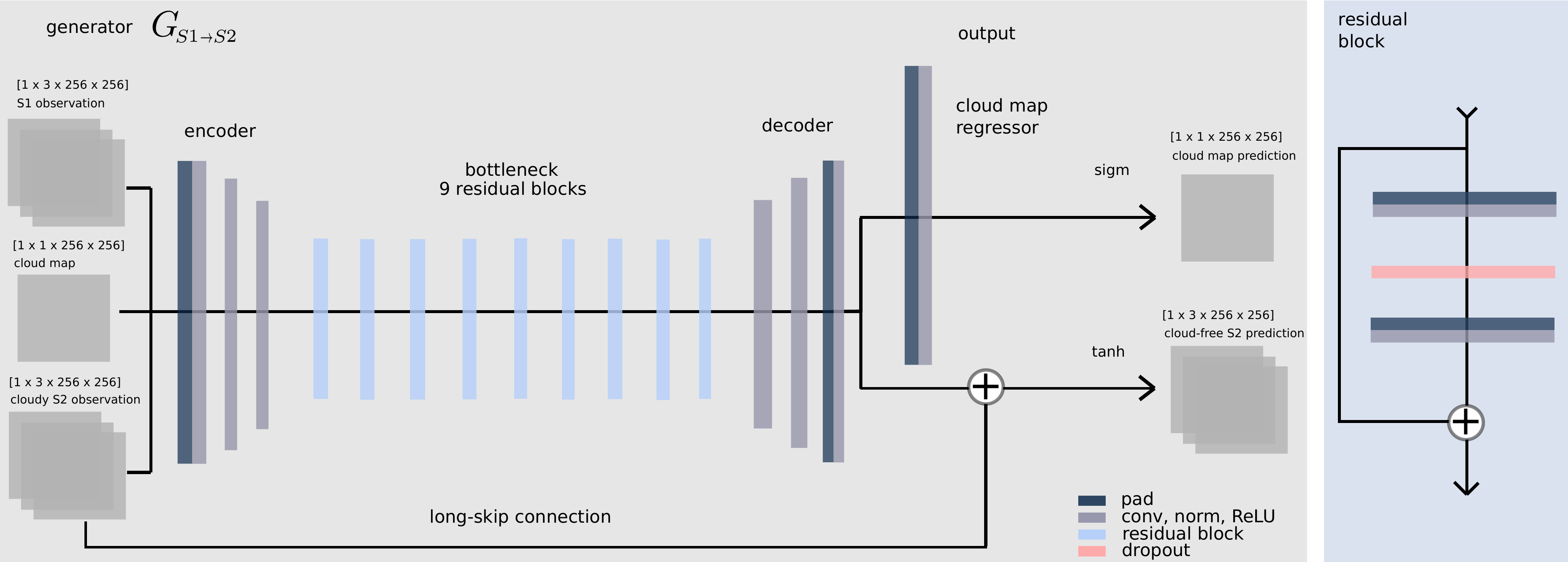}
    \caption{Detailed architecture of the generator $G_{S1 \rightarrow S2}$ of Fig. \ref{fig:architecture}. The generator receives $S1$, $m$ and $S2$ as input, the latter of which is long-skip forwarded and modified by the learned residual map $S2_{res}$. The result is passed via a non-linearity as input to the next network, or treated as output. In parallel, $S2_{res}$ is regressing $m$ to enforce sparseness of the residual map.}
    \label{fig:generator}
\end{figure*}

\subsection{Architecture} \label{sub:architecture}
The model proposed in this work follows the architecture of cycle-consistent GAN \cite{Zhu_Park_Isola_Efros_2017}, i.e. we use two generative networks $G_{S1 \rightarrow S2}$ and $G_{S2 \rightarrow S1}$ that translate images from the source   domain of $S1$ to the target domain of $S2$ and vice versa. Distribution $\dot{S1}$ (or $\dot{S2}$) denotes the target when the generator performs a within-domain identity mapping, preserving the input image's sensor characteristics. For each domain there exists an associated discriminator network, denoted as $D_ {S1}$ and $D_{S2}$ respectively, classifying whether a given image is a sample from the domain's true distribution S1 (or S2) or from the synthesized distribution $\hat{S1}$ (or $\hat{S2}$). An overview of our model ensemble is given in Fig. \ref{fig:architecture}. While we keep the network $G_{S2 \rightarrow S1}$ as in the original work, we apply spectral normalization \cite{Miyato_Kataoka_Koyama_Yoshida_2018} to both discriminators and make adjustments as follows: 
$G_{S1 \rightarrow S2}$ receives an image from domain $S1$ as input and is additionally conditioned on the corresponding cloudy image from $S2$ as well the cloud probability mask $m$. For our cloud-removal network we keep the encoder-decoder architecture of the generator but add a long-skip connection \cite{He_Zhang_Ren_Sun_2016}
such that the output is given by 
$$\hat{S2} = G_{S1 \rightarrow S2}(\cdot) = tanh(S2 + S2_{res})$$

where $S2_{res}$ denotes the residual mapping learned by the generator. To demodulate the effects of the output non-linearity on the long-skipped pixels the inverse hyperbolic tangent is applied to the cloudy input image from $S2$ before the residual mapping. 
Furthermore, we insert a regression layer taking the residual maps $S2_{res}$ as input and returning a prediction $\hat{m}$ of the cloud map $m$. The purpose of the regressor is to enforce a meaningful relation between the learned $S2_{res}$ and the conditioning $m$, making the residual maps sparse. Here, sparseness refers to the residual maps being (close to) zero over non-cloudy areas, as opposed to having widespread small values which would indicate many unneeded changes made to cloud-free pixels. We enforce sparseness of the residual maps by formulating an L1 loss on the cloud mask regression as defined in section \ref{sub:cloudmap}. The loss term effectively acts as a regularizer on changes made to non-cloudy areas, penalizing unnecessary adjustments. The regression layer consists of a [3$\times$3] convolutional kernel mapping the generated three-dimensional image to a single-channel map and thus adds little to the overall number of learnable parameters. The architecture of generator $G_{S1 \rightarrow S2}$ is depicted in Fig. \ref{fig:generator}, and details on its parametrization are provided in Table \ref{table:generator}.
Discriminator $D_ {S2}$ is as well conditioned on the cloud probability maps $m$.
Importantly, we forward the (unpaired) non-cloudy optical images to the discriminator $D_ {S2}$, which learns the non-cloudy patch-wise statistics and thus implicitly forces $G_{S1 \rightarrow S2}$ to synthesize cloud-free images. In sum, our main contribution with respect to architectural changes is two-fold: First, we adjusted the generator predicting cloud-free optical images to learn a residual mapping by introducing a long-skip connection forwarding optical information, removing the previous need to reconstruct (even cloud free) pixels from scratch. Second, our generator learns to constrain modifications to cloud-covered pixels while keeping clear areas unchanged, which is encouraged by introducing a novel layer regressing the cloud coverage map by the learned residual map. 

% \usepackage{multirow}
\begin{table*}[h!tb] 
\centering
\begin{tabular}{l|ll|l|l}
encoder                   & \multicolumn{2}{l|}{bottleneck}    & decoder                                    & output                                      \\ \hline
R(N(C(3 $\times$ 3, 64, 1, 1)))  &     & R(N(C(3 $\times$ 3, 256, 1, 1)))    & R(N(T(3 $\times$ 3, 256, 2, 1)))                  & sigmoid(C(3 $\times$ 3, 1, 1, 1))                     \\ \cline{5-5} 
R(N(C(3 $\times$ 3, 128, 2, 1))) & 9 $\times$ & dropout(0.5)                      & \multirow{2}{*}{R(N(T(3 $\times$ 3, 128, 2, 1)))} & \multirow{2}{*}{tanh(C(3 $\times$ 3, 3, 1, 1))} \\
R(N(C(3 $\times$ 3, 256, 2, 1))) &     & R(N(C(3 $\times$ 3, 256, 1, 1))) &                                            &                                            
\end{tabular}
\caption{Architecture of our generator $G_{S1 \rightarrow S2}$. The architecture is divided into 4 components as illustrated in Fig. \ref{fig:generator}, information flow is from left to right across components and top to bottom within components. Symbols: R (ReLU), N (instance normalization), C (convolution), T (transposed convolution). For (transposed) convolution the parametrization is (kernel height $\times$ kernel width, number of filters, stride, padding size). The architecture of generator $G_{S2 \rightarrow S1}$ is as the 9-ResNet block generator in \cite{Zhu_Park_Isola_Efros_2017} and the two discriminators are kept as the PatchGAN discriminators in \cite{Zhu_Park_Isola_Efros_2017}.}
\label{table:generator}
\end{table*}

\subsection{Losses} \label{sub:losses}
We adjust the losses such that regions regressed as cloud-free in map $m$ remain untouched while cloudy areas are recovered given the information from domain $S1$. The losses minimized by the generators are
\begin{alignat*}{2}
&\mathcal{L}_{adv} &&= (D_{S1}(\hat{S1})-1)^2 + (D_{S2}(\hat{S2})-1)^2  \\
&\mathcal{L}_{cyc} &&= ||m \cdot (S1-\breve{S1})||_1 + ||(1-m) \cdot (S2-\breve{S2})||_1 \\
&\mathcal{L}_{idt} &&= ||m \cdot (S1-\dot{S1})||_1 + ||m \cdot (S2-\dot{S2})||_1 \\
&\mathcal{L}_{aux} &&= ||(1-m) \cdot (m-\hat{m})||_1 \\
&\mathcal{L}_{all} &&= \lambda_{adv} \mathcal{L}_{adv} + \lambda_{cyc} \mathcal{L}_{cyc} + \lambda_{idt} \mathcal{L}_{idt} + \lambda_{aux} \mathcal{L}_{aux},
\end{alignat*}
where $\lambda_{adv}=5.0$, $\lambda_{cyc}=10.0$, $\lambda_{idt}=1.0$ and $\lambda_{aux}=10.0$ are hyper-parameters to linearly combine the individual losses within $\mathcal{L}_{all}$. The loss weightings are set similar to those in \cite{Zhu_Park_Isola_Efros_2017}, with minor adjustments made manually. 
$\mathcal{L}_{adv}$ is the adversarial loss originally proposed in LSGAN \cite{Mao_Li_Xie_Lau_Wang_Smolley_2017}, implementing a least-squares error function on the classifications of the discriminators $D_{S1}$ and $D_{S2}$. $\mathcal{L}_{cyc}$ and  $\mathcal{L}_{idt}$ are as introduced in \cite{Zhu_Park_Isola_Efros_2017} but weighted pixel-wise with the cloud map $m$. The purpose of the cycle-consistent loss $\mathcal{L}_{cyc}$ is to regularizing the mapping $S1 \rightarrow S2$ by requiring $S2 \rightarrow S1$ being able to reconstruct the original input again (likewise for the direction $S2 \rightarrow S1 \rightarrow S2$), constraining the potential mappings between both domains. The idea behind $\mathcal{L}_{idt}$ is to motivate generators to perform an identity mapping and limit unneeded changes in case the provided input is a sample of the target domain. $\mathcal{L}_{aux}$ is the loss associated with the cloud map regression in $G_{S1 \rightarrow S2}$, introduced to enforce sparseness of the learned residual feature maps $S2_{res}$ such that the non-cloudy pixels of $S2$ experience little to no adjustments. Our modified generator architecture, the usage of probabilistic cloud maps and the adjusted losses are showcased in context of a cycle-consistent GAN ensemble, but we remark that they may as well be used within alternative models such as conditional GAN \cite{Mirza_Osindero_2014} or ResNet architectures \cite{He_Zhang_Ren_Sun_2016}.

\section{Experiments and Analysis}
\label{sec:experiments}

\subsection{Data} \label{data}

\begin{figure}[h!tb]
  \centering
  \begin{subfigure}[b]{0.49\linewidth}
    \includegraphics[width=\linewidth]{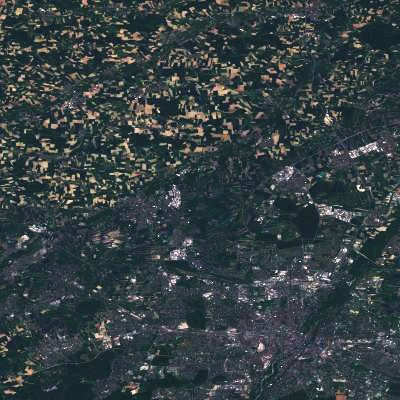}
  \end{subfigure}
    \begin{subfigure}[b]{0.49\linewidth}
    \includegraphics[width=\linewidth]{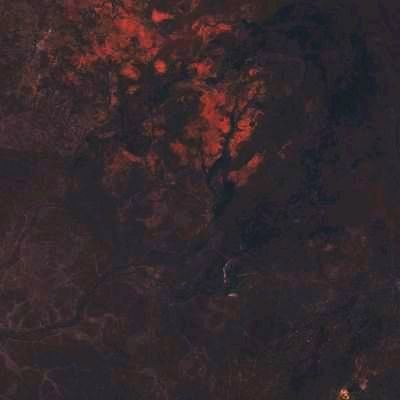}
  \end{subfigure}
  \label{fig:coffee}
  \begin{subfigure}[b]{0.49\linewidth}
    \includegraphics[width=\linewidth]{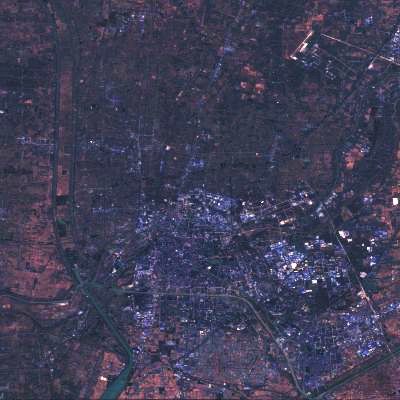}
  \end{subfigure}
    \begin{subfigure}[b]{0.49\linewidth}
    \includegraphics[width=\linewidth]{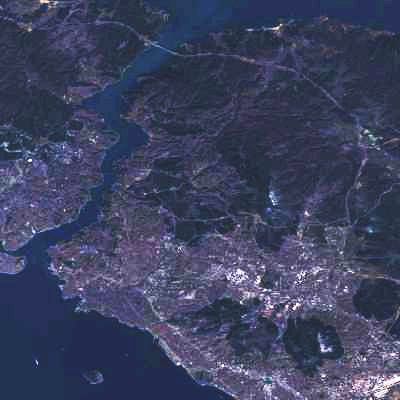}
  \end{subfigure}
  \caption{Cloudless S2 imagery of four exemplary ROI, illustrating the diversity of SEN12MS-CR. The four different scenes are of four different meteorological seasons from the test split of the data set. On average a ROI is split into over 700 patch samples, each observation of size $256 \times 256 \: px^2$.}
  \label{fig:diverseROI}
\end{figure}

To conduct our experiments we gather a novel large-scale data set called SEN12MS-CR for cloud removal. For this purpose, we build upon the openly available SEN12MS data set \cite{Schmitt_Hughes_Qiu_Zhu_2019} of globally sampled co-registered $S1$ plus cloud-free $S2$ patches and complement the data set with co-registered cloudy images close in time to the original observations. SEN12MS-CR consists of 169 non-overlapping regions of interest (ROI) evenly distributed over all continents and meteorological seasons. The ROI have an average size of approximately $52 \times 40 \: km^2$ ground coverage, corresponding to complete-scene images of about $5200 \times 4000 \: px^2$. Each complete-scene image is checked manually to ensure freedom of noise and artifacts. The cloud-free optical images of four exemplary ROI  observed in four different meteorological seasons are depicted in Fig. \ref{fig:diverseROI} to highlight the heterogeneity of landcover captured by SEN12MS-CR. Each scene in the data set is subsequently translated into Universal Transverse Mercator coordinate system and then partitioned into patches of size $256 \times 256 \: px^2$ with a spatial overlap of $50 \%$ between neighboring patches, yielding an average of over 700 patches per ROI. Each patch consists of a triplet of orthorectified, geo-referenced cloudy and cloud-free 13-band multi-spectral Sentinel-2 images, as well as the correspondent Sentinel-1 image (see Fig. \ref{fig:teaser} for examples of SAR, cloud-free and cloudy optical patch triplets).
Paired images of the three modalities were acquired within the same meteorological season to limit surface changes.
The Sentinel-2 data is from the Level-1C top-of-atmosphere reflectance product. Finally, each patch triples is automatically controlled for potential imaging artifacts and exclusively artifact-free patches are preserved to constitute the final cleaned-up version of SEN12MS-CR.

\begin{figure}[h!tb]
    \includegraphics[width=\linewidth]{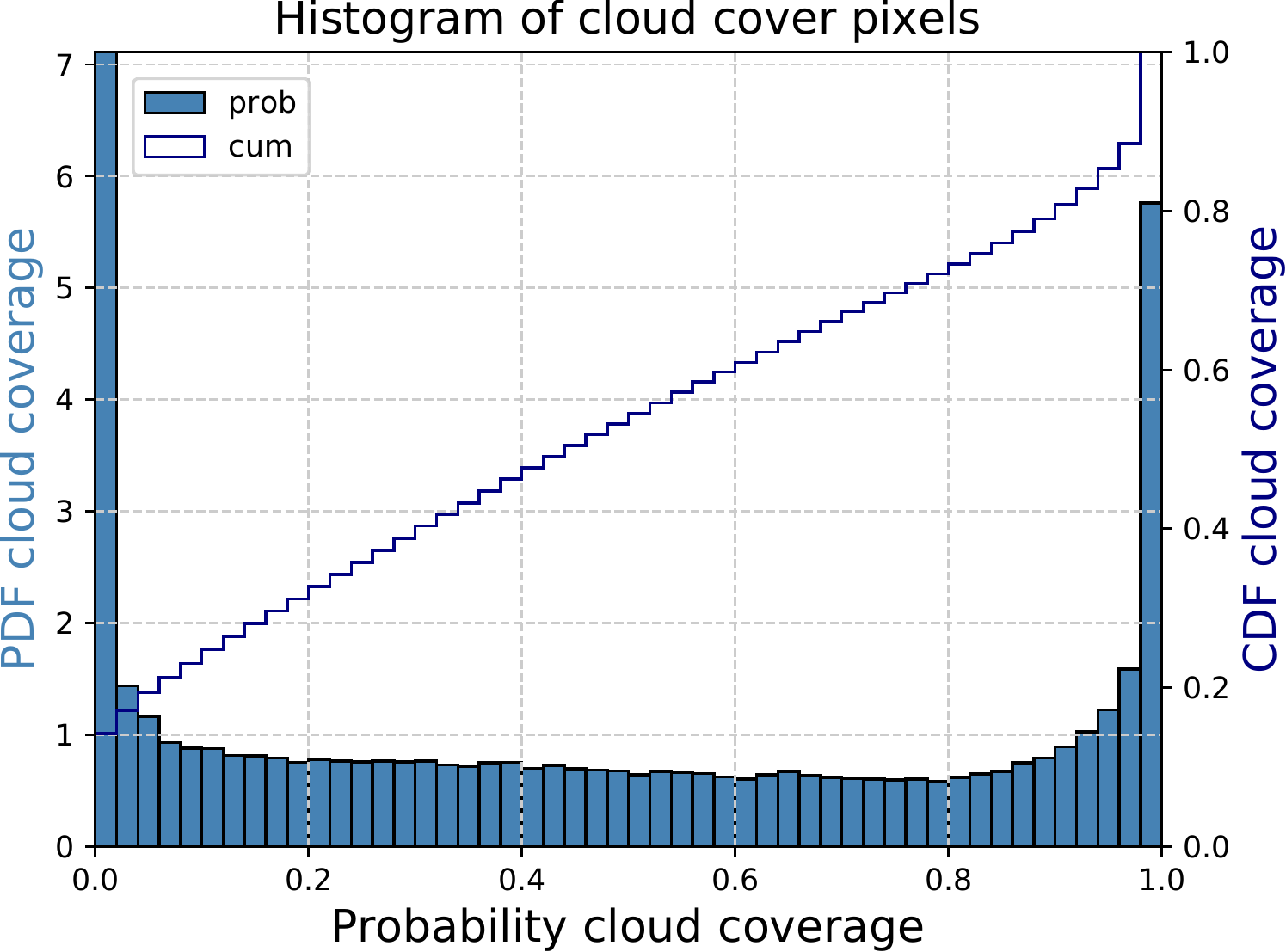}
    \caption{Statistics of cloud coverage of SEN12MS-CR. On average, approximately $50 \%$ of occlusion is observed. The empirical distribution of cloud coverage is relatively uniform and ranges from cloud-free views to total occlusion.}
    \label{fig:cloud_coverage}
\end{figure}

Evaluating the cloudiness of each patch with the algorithm of \cite{Zupanc} as described in section \ref{sub:cloudmap} yields a mean cloud coverage of circa $47.93 \pm 36.08$ percent, i.e. about half of all optical images' information is affected by clouds and the amount of coverage varies considerably. This amount of coverage is notably close to the approximately $55$ percent of global cloud fraction over land that have previously been observed empirically \cite{King_Platnick_Menzel_Ackerman_Hubanks_2013}.
The distribution of cloud coverage is shown in Figure \ref{fig:cloud_coverage} and is relatively uniform over the entire domain, with slightly more samples showing (almost) no clouds or being entirely cloud-covered. Note that the computed cloud probability masks are not used to filter any observations or actively guide the data set creation in any manner, they are solely used post-hoc to quantify the distribution of cloudiness. For the sake of comparability across models in our experiments and for further studies, we define a train split and a split of hold-out data which is reserved for the purpose of testing. The train split consists of 114325 patches sampled uniformly across all continents and seasons and is open to be entirely used for training or in parts for training and validating. The test split consists of 7893 geospatially separated images sampled from ten different ROI distributed across all continents and meteorological seasons, capturing a heterogeneous subset of data.

\subsection{Experiments and Results}

A total of three experiments are conducted. First, we train our network and extend it by adding supervised losses for the model to benefit of paired non-cloudy and cloudy optical observations in our data set at training time. We systematically vary the amount of available supervision to investigate its effects on model performance. Second we evaluate it against a set of baseline models. Third, we re-train the architectures from the previous experiment on synthetic data of generated cloudy observations and evaluate them on real data in order to quantify to which extent models trained on simulated data are capable to generalize to real-world scenarios. To our knowledge, neither of these experiments have previously been conducted in depth. All experiments were conducted on a machine of 8 Intel(R) Core(TM) i7-8700 CPU @ 3.20GHz processors, 16GB of DIMM DDR4 Synchronous 2667 MHz RAM and an NVIDIA GeForce RTX 2080, running Ubuntu 18.04. Computation clock time for the training procedure may vary according to the overall task load but is estimated to be about 7 days for model ours-0 and about 10-12 days for model ours-100.

\subsubsection{Metrics to quantify the goodness of cloud removal}
In order to evaluate model performances quantitatively, we utilize the recently developed metrics of improved precision and recall \cite{Kynkaanniemi_Karras_Laine_Lehtinen_Aila_2019}, proposed in the context of generative modeling and improving on previous metrics such as Inception score or Frechét Inception distance \cite{Salimans_Goodfellow_Zaremba_Cheung_Radford_Chen_Chen, Heusel_Ramsauer_Unterthiner_Nessler_Hochreiter}. Improved precision and recall are measures of goodness quantifying similarities between two sets of images in a high-dimensional feature embedding space. Precision is a metric of sample quality, assessing the fraction of generated images that are plausible in the context of the target data distribution. In our context, a generated image is plausible if its high-dimensional feature embedding is sufficiently close to the high-dimensional feature embedding of a cloud-free target image. The distance between both embeddings is sufficiently small if there’s no fixed number of neighbors closer to the target embedding than the query embedding is. For the formalities behind this metric and a motivation of the chosen parametrization please see Appendix \ref{appendix:b}. Recall measures the diversity in generated images and the extent to which the distribution of target data is covered. Analogous to the metric of precision, a target image is recalled if its high-dimensional feature embedding is sufficiently close to the high-dimensional feature embedding of a generated cloud-free image. Note that this allows to interpret recall as a measure of generated image diversity as the metric can score high only if the generated samples are spread out in the feature embedding’s space and provide sufficient coverage of the distribution of target images, capturing the heterogeneity of the target images. To summarize, in the context of our data set of section \ref{data}, precision specifies the closeness of cloud-recovered information to its cloud-free counterpart, whereas recall captures how well the de-clouded images capture the heterogeneity of the test data (e.g. its diversity in land-cover, seasonality, etc).

While we emphasize the benefit of both measures to disentangle image quality and image heterogeneity we also define the F1 score as
$$F1(X,Y) = 2 \cdot \frac{PR(X,Y) \cdot RC(X,Y)}{PR(X,Y) + RC(X,Y)}$$
where X, Y are sets of images to be compared and PR, RC denote the functions of precision and recall, respectively.
In contrast to the first two experiments, the generation of synthetic data in the third experiment guarantees a one-to-one pixelwise correspondence between cloudy and ground-truth cloud free images (i.e. perfect co-registration, no atmospheric disturbances other than the simulated noise, control for no landcover and daylight changes between both observations), ensuring that pixel-wise metrics are well-defined. Therefore, complementary to the previous measures of goodness, we additionally assess performances on synthetic data in the third experiment by means of mean absolute error (MAE), root mean squares error (RMSE), Peak Signal-to-Noise Ratio (PSNR), structural similarity (SSIM) \cite{Wang_Bovik_Sheikh_Simoncelli_2004} and Spectral Angle Mapper (SAM) \cite{kruse1993spectral} as given by 
$$MAE(x,y) = \frac{1}{C \cdot H \cdot W} \sum_{c=h=w=1}^{C, H, W} |x_{c, h, w}-y_{c, h, w}|$$ 
$$RMSE(x,y) = \sqrt{\frac{1}{C \cdot H \cdot W} \sum_{c=h=w=1}^{C, H, W} (x_{c, h, w}-y_{c, h, w})^2}$$ 
$$PSNR(x,y) = 20 \cdot log_{10} \left( \frac{1}{RMSE(x,y)} \right)$$ 
$$SSIM(x,y) = \frac{(2 \mu_x\mu_y + \epsilon_1)(2 \sigma_{xy} + \epsilon_2)}{(\mu_x+\mu_y+\epsilon_1)(\sigma_x + \sigma_y + \epsilon_2)}$$
\small $$SAM(x,y) = cos^{-1} \left( \frac{\sum^{C, H, W}_{c=h=w=1} x_{c, h, w} \cdot y_{c, h, w}}{\sqrt{\sum^{C, H, W}_{c=h=w=1} x_{c, h, w}^2 \cdot \sum^{C, H, W}_{c=h=w=1} y_{c, h, w}^2}} \right)$$
\normalsize
where $x,y$ are images to be compared with pixel-values $x_{c, h, w}, y_{c, h, w} \in [0,1]$, dimensions $C=3$, $H=W=256$, means $\mu_x, \mu_y$, standard deviations $\sigma_x, \sigma_y$, covariance  $\sigma_{xy}$ and small numbers  $\epsilon_1$, $\epsilon_2$ to stabilize the computation. MAE and RMSE both are pixel-level metrics quantifying the mean deviation between target and predicted images in absolute terms and units of the measure of interest, respectively. PSNR is an image-wise metric to measure how good of a reconstruction in terms of signal-to-noise a recovered image is to a clear target image. SSIM is a second image-wise metric quantifying structural differences between the target and predicted images. It is designed to capture perceived change in structural information between two given images, as well as differences in luminance and contrast \cite{Wang_Bovik_Sheikh_Simoncelli_2004}. The SAM metric is an image-wise measure quantifying the spectral angle between two images, measuring their similarity in terms of rotations in the space of spectral bands \cite{kruse1993spectral}. Further technical information with respect to the metrics utilized in our experiments to quantify goodness of predictions is provided in appendix \ref{appendix:b}.

\subsubsection{Quantifying the benefits of paired data}
First, we train the architecture described in section \ref{sec:methods} without using any pixel-wise correspondences, as in a manner conventional for cycle-consistent GAN. For our generative model we consider the VV and VH channels of images from the S1 domain and add a third mean(VV,VH) channel to satisfy the dimension-preservation requirement of cycle-consistent architectures. For images from the S2 domain all multi-spectral information is used when computing cloud probability maps while the S1-S2 mapping uses exclusively the three RGB channels.
All images are value-clipped and rescaled to contain values within $[-1,1]$, while the cloud probability map values are within $[0,1]$. Value-clipping is within ranges $[-25;0]$ and $[0;10000]$ for S1 and S2, respectively.
Notably, before training we perform an image-wise shuffling of the optical data of paired cloudy and cloud-free observations to remove the pixelwise correspondences satisfied when cloudy and cloud-free patches would be available as sorted tuples. That is, the optical cloudy and non-cloudy patches presented at one training step may be no longer strictly aligned or could reflect differences in landcover and atmosphere, reflecting practical challenges commonly encountered when gathering data for remote sensing applications. We train our network on a 10000 images multi-region subset of the training split introduced in section \ref{data}. Network weights $w$ are initialized by sampling from a Gaussian distribution $w \sim \mathcal{N}(\mu = 0, \sigma^2 = 0.02)$. The optimizer as well as the hyperparameters for the optimizer and loss weightings are set as in \cite{Zhu_Park_Isola_Efros_2017}: We use ADAM with an initial learning rate $\epsilon_{lr} = 0.0002$, momentum parameters $\beta = (0.5, 0.999)$ for computing sliding averages of the gradients as well as their squares and a small constant of $10^{-8}$ added to the denominator to ensure numerical stability of the optimizer. Instance normalization \cite{Ulyanov_Vedaldi_Lempitsky_2017} is applied to the generators as in the original architecture \cite{Zhu_Park_Isola_Efros_2017}, with adjustments detailed in Fig. \ref{fig:generator} and Table \ref{table:generator}. Spectral normalization \cite{Miyato_Kataoka_Koyama_Yoshida_2018} is applied to the discriminators as in \cite{mo2019instagan} in order to prevent mode collapse during training \cite{goodfellow2016nips}.
The networks are trained for $n_{iter} = 50$ epochs at the initial learning rate of $\epsilon_{lr}$, then for another $n_{decay} = 25$ epochs with a multiplicative learning rate decay given by $lr_{decay} (n_{current})= 1.0 - max(0, 1 + n_{current}- n_{iter}) / (n_{decay} + 1)$, where $n_{current}$ denotes the current epoch number. The gentle learning rate decay over a long period of epochs serves to ensure a well-behaved optimization process during training \cite{Zhu_Park_Isola_Efros_2017, goodfellow2016nips}.
All our generator networks are trained on center-cropped $200 \times 200 \: px^2$ patches but tested on full-sized $256 \times 256 \: px^2$ patches of the hold-out split, as the generator architecture is fully convolutional. As proposed in \cite{shrivastava2017learning} and implemented in \cite{ Zhu_Park_Isola_Efros_2017} we maintain two pools of the last 50 generated images to update the discriminators with a random sample from the respective image buffers such that oscillations during training are reduced \cite{Zhu_Park_Isola_Efros_2017, goodfellow2016nips}. Representative qualitative outcomes are depicted in Fig. \ref{fig:teaser}. The results highlight that our model can reconstruct cloud-covered areas while preserving information that is not obscured. A quantitative evaluation of the described model (ours-0) is given in Table \ref{tab:paired}. 

\begin{table}[h!tb]
\centering
\resizebox{0.35\textwidth}{!}{%
\begin{tabular}{llll}
\hline
\% paired & \multicolumn{1}{c}{precision} & recall           & \multicolumn{1}{r}{F1 score} \\ \hline
0 (ours-0)        & 0.560                         & 0.491            & 0.523                        \\
10        & 0.559                         & 0.499            & 0.527                        \\
20        & 0.560                         & 0.506            & 0.532                        \\
50        & 0.562                         & 0.528            & 0.544                        \\
100 (ours-100)      & $\mathbf{0.564}$              & $\mathbf{0.551}$ & $\mathbf{0.557}$             \\ \hline
\end{tabular}%
}
    \caption{Effect of percentage of paired trained data on performance of cloud removal model. The more paired training data is available, the better the resulting performances.} 
    \label{tab:paired}
\end{table}

Second, we re-train the model as described above, but on paired cloudy-cloudless optical observations in order to assess the benefits of paired training data, as provided by our data set. To let the cycle-consistent architecture described in section \ref{sec:methods} benefit of paired training data, we combine the losses defined in section \ref{sub:losses} with cost functions defined on paired images: First, a pixel-wise L1 loss penalizing prediction errors between generated and paired target images as in \cite{isola2017image}. Second, perceptual losses for features and style \cite{Johnson_Alahi_Fei-Fei_2016}, as evaluated on the features extracted at ReLU layers 11, 20 and 29 of an auxiliary pre-trained VGG16 network \cite{Simonyan_Zisserman_2014}. We re-train our network with these losses and systematically vary the percent of paired cloudy and cloud-free optical data available. The paired patches are equally spaced across the training split at the beginning of the training procedure and patch pairings are fixed across epochs. During training, the presentation of paired and unpaired samples occurs in a random order. Table \ref{tab:paired} shows the different models' performances. The base model trained on unpaired data (ours-0) performs worst while the model fully trained on paired samples (ours-100) achieves the best performances. In general, the more paired samples are available the better the model performs.

\comment{
\begin{table}[h!tb]
\centering
\resizebox{0.4\textwidth}{!}{%
\begin{tabular}{ccccc}
\hline
\multicolumn{2}{c}{model}           & precision        & recall           & F1 score         \\ \hline
\multirow{2}{*}{\begin{tabular}[c]{@{}c@{}}S1\end{tabular}} & VV  & 0.000             & 0.001             & 0.001             \\ 
                               & VH & 0.012             & 0.017             & 0.014             \\ 
\multicolumn{2}{c}{S2 cloudy}          & 0.161            & $\mathbf{0.705}$ & 0.267            \\ 
\multicolumn{2}{c}{Grohnfeldt \cite{Grohnfeldt_Schmitt_Zhu_2018}}      & 0.047             & 0.003             & 0.005             \\ 
\multicolumn{2}{c}{Bermudez \cite{Bermudez_Happ_Oliveira_Feitosa_2018}}        & TBC             & TBC             & TBC             \\ 
\multirow{2}{*}{\begin{tabular}[c]{@{}c@{}}Fuentes-\\ Reyes \cite{Fuentes_Reyes_Auer_Merkle_Henry_Schmitt_2019} \end{tabular}} & RGB  & 0.000             & 0.103             & 0.000               \\ 
                               & mono & 0.000             & 0.210             & 0.000             \\ 
\multicolumn{2}{c}{Anonymized \cite{Meraner}}         & 0.462            & 0.567            & 0.510            \\ 
\multicolumn{2}{c}{ours-0}          & 0.560            & 0.491            & 0.523            \\ 
\multicolumn{2}{c}{ours-100}        & $\mathbf{0.564}$ & 0.551            & $\mathbf{0.557}$ \\ \hline
\end{tabular}
}
    \caption{Cloud-removal performance of baseline methods and our models on test split of SEN12MS-CR.}
    \label{tab:results}
\end{table}
}

\begin{table}[h!tb]
\centering
\resizebox{0.4\textwidth}{!}{%
\begin{tabular}{ccccc}
\hline
\multicolumn{2}{c}{model}           & precision        & recall           & F1 score         \\ \hline
\multirow{2}{*}{\begin{tabular}[c]{@{}c@{}}S1\end{tabular}} & VV  & 0.000             & 0.001             & 0.001             \\ 
                               & VH & 0.012             & 0.017             & 0.014             \\ 
\multicolumn{2}{c}{S2 cloudy}          & 0.161            & $\mathbf{0.705}$ & 0.267            \\ \hline
\multicolumn{2}{c}{ours-0 (no $m$)}      & 0.181             & 0.572             & 0.279             \\ 
\multicolumn{2}{c}{ours-100 (no $m$)}      & 0.232             & 0.535             & 0.323             \\ 
\multicolumn{2}{c}{ours-0}          & 0.560            & 0.491            & 0.523            \\ 
\multicolumn{2}{c}{ours-100}        & $\mathbf{0.564}$ & 0.551            & $\mathbf{0.557}$ \\ \hline
\end{tabular}
}
    \caption{Cloud-removal performance of baseline methods and our models on test split of SEN12MS-CR. Rows S1 VV an VH refer to the raw S1 image, channels VV and VH respectively, compared to the gray-scale cloud-free S2 image. S2 cloudy refers to the raw cloudy S2 image, compared to the RGB cloud-free S2 image. All models metrics beat the lower-bound performances established by the raw data, except on the recall metric. The full models perform better than the ablation models without the cloud-sensitive loss and cloud probability masks. Model ours-100 performs best in terms of precision and F1 score. Note that the results depict a pronounced trade-off between precision and recall, analysed in detail in \cite{Kynkaanniemi_Karras_Laine_Lehtinen_Aila_2019}.}
    \label{tab:results}
\end{table}

\subsubsection{Model ablation experiment}

To put the results of the previous experiment into perspective and further evaluate the factors benefiting robust reconstruction of cloud-covered information we conduct an ablation study. Specifically, we investigate the effectiveness of the novel cloud detection mechanism explained in section \ref{sub:cloudmap} and the local cloud-sensitive loss introduced in section \ref{sub:losses}. For this purpose, we re-train the model ours-0 as described in section \ref{sec:methods}, but omit the cloud-sensitive terms by fixating the values of all pixels in the cloud probability masks $m$ to $1.0$. The effect of this is that the ablated model is no longer encouraged to minimize changes to areas free of cloud coverage, thus potentially resulting in unneeded changes. As additional baselines, we evaluate the goodness of simply using the S1 observations (VV- or VH-polarized) as well as cloud-covered S2 images as predictions and comparing against their cloud-free counterparts. Table \ref{tab:results} reports the de-clouding performance of baseline models and our models (0 and 100 percent paired data from Table \ref{tab:paired}). Our network of 100 \% paired data performs best in terms of precision and F1 score. The raw S1 and S2 observations perform relatively poorly, except for the cloudy optical images scoring high on image diversity due to random cloud coverage. While it may be useful to consider the raw data as baselines it is necessary to keep in mind that modalities like SAR may be at a disadvantage when directly comparing against the cloud-free optical target images.

\begin{table}[]
\centering
\begin{tabular}{ccclcl}
\hline
\multicolumn{2}{c}{model}          & \multicolumn{2}{c}{ours-0}         & \multicolumn{2}{c}{ours-100}       \\ \hline
\multicolumn{2}{c|}{metric}        & \multicolumn{1}{l}{Perlin} & copy  & \multicolumn{1}{l}{Perlin} & copy  \\ \hline
\multicolumn{2}{c}{MAE}           & 0.045                      & 0.023 & 0.041                      & 0.017 \\
\multicolumn{2}{c}{RMSE}           & 0.067                      & 0.031 & 0.059                      & 0.023 \\
\multicolumn{2}{c}{PSNR}           & 24.75                      & 34.034 & 25.775                      & 35.802 \\
\multicolumn{2}{c}{SSIM}           & 0.803                      & 0.882 & 0.824                      & 0.904 \\
\multicolumn{2}{c}{SAM}           & 27.527                      & 10.626 & 26.013                      & 9.936 \\
\multirow{2}{*}{precision} & synth & 0.155                      & 0.693 & 0.239                      & 0.692 \\
                           & real  & 0.115                      & 0.425 & 0.168                      & 0.458 \\
\multirow{2}{*}{recall}    & synth & 0.781                      & 0.851 & 0.800                      & 0.856 \\
                           & real  & 0.624                      & 0.611 & 0.592                      & 0.586 \\
\multirow{2}{*}{F1}        & synth & 0.258                      & 0.764 & 0.368                      & 0.766 \\
                           & real  & 0.194                      & 0.501 & 0.262                      & 0.514 \\ \hline
\end{tabular}%
    \caption{Cloud-removal performance of models ours-0 and ours-100 from Table \ref{tab:results}, re-trained on synthetic cloud data (either Perlin-simulated or copy-pasted) and tested on synthetic as well as real data. Both models, when trained on synthetic data, perform much better on synthetic test data than on real test data. Importantly, the test performance of models trained on synthetic and tested on real data is considerably poorer than that of the same architectures trained on real data (reported in Table \ref{tab:results}).
    }
    \label{tab:synthetic}
\end{table}

\subsubsection{Assessing the goodness of synthetic data} \label{subsub:synthetic}

\begin{figure}[h!tb]
  \centering
  \begin{subfigure}[b]{0.25\linewidth}
    \includegraphics[width=\linewidth]{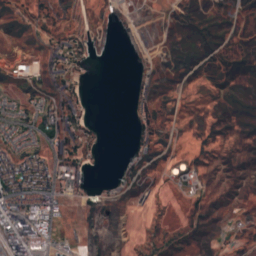}
  \end{subfigure}
  \begin{subfigure}[b]{0.25\linewidth}
    \includegraphics[width=\linewidth]{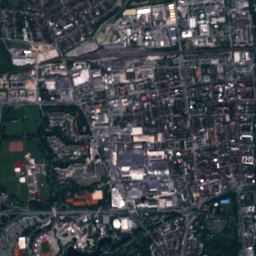}
  \end{subfigure}
    \begin{subfigure}[b]{0.25\linewidth}
    \includegraphics[width=\linewidth]{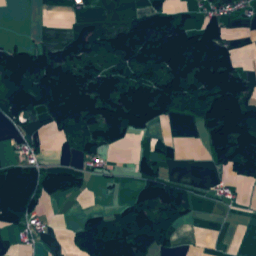}
  \end{subfigure}
  
  \begin{subfigure}[b]{0.25\linewidth}
    \includegraphics[width=\linewidth]{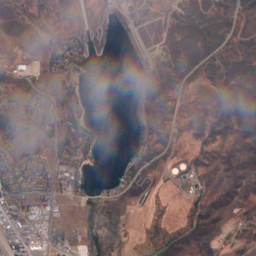}
  \end{subfigure}
  \begin{subfigure}[b]{0.25\linewidth}
    \includegraphics[width=\linewidth]{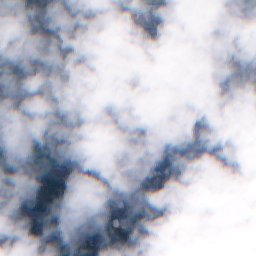}
  \end{subfigure}
    \begin{subfigure}[b]{0.25\linewidth}
    \includegraphics[width=\linewidth]{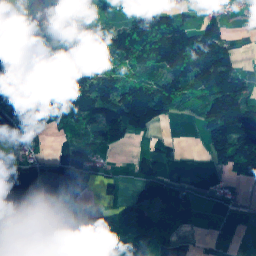}
  \end{subfigure}
  
   \begin{subfigure}[b]{0.25\linewidth}
    \includegraphics[width=\linewidth]{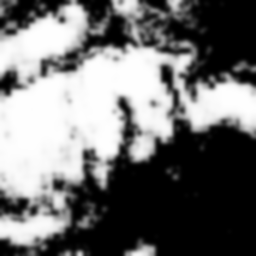}
  \end{subfigure}
  \begin{subfigure}[b]{0.25\linewidth}
    \includegraphics[width=\linewidth]{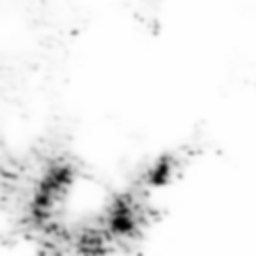}
  \end{subfigure}
    \begin{subfigure}[b]{0.25\linewidth}
    \includegraphics[width=\linewidth]{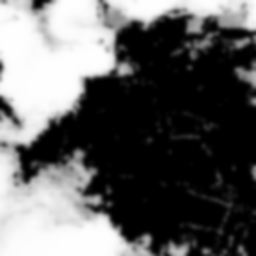}
  \end{subfigure}
  
  \begin{subfigure}[b]{0.25\linewidth}
    \includegraphics[width=\linewidth]{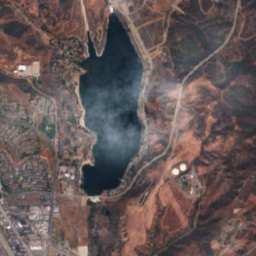}
  \end{subfigure}
  \begin{subfigure}[b]{0.25\linewidth}
    \includegraphics[width=\linewidth]{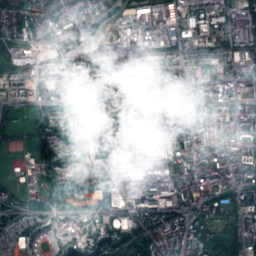}
  \end{subfigure}
    \begin{subfigure}[b]{0.25\linewidth}
    \includegraphics[width=\linewidth]{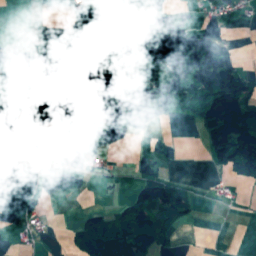}
  \end{subfigure}
  
   \begin{subfigure}[b]{0.25\linewidth}
    \includegraphics[width=\linewidth]{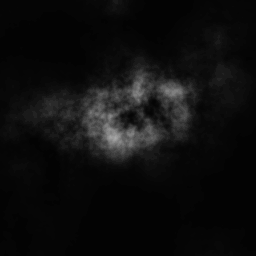}
  \end{subfigure}
  \begin{subfigure}[b]{0.25\linewidth}
    \includegraphics[width=\linewidth]{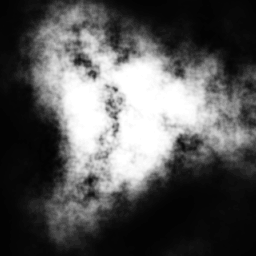}
  \end{subfigure}
    \begin{subfigure}[b]{0.25\linewidth}
    \includegraphics[width=\linewidth]{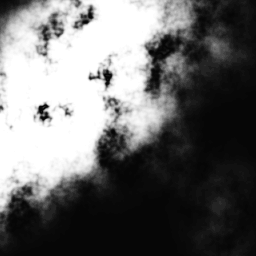}
  \end{subfigure}
  
   \begin{subfigure}[b]{0.25\linewidth}
    \includegraphics[width=\linewidth]{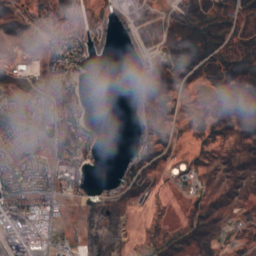}
  \end{subfigure}
  \begin{subfigure}[b]{0.25\linewidth}
    \includegraphics[width=\linewidth]{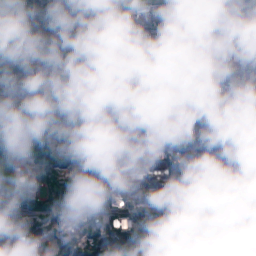}
  \end{subfigure}
    \begin{subfigure}[b]{0.25\linewidth}
    \includegraphics[width=\linewidth]{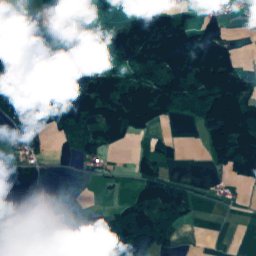}
  \end{subfigure}
  \caption{Exemplary cloud-free, real cloudy and generated cloudy optical observations. Rows: Cloud-free S2 data (plotted in RGB), real cloudy S2 data, real cloud coverage maps (same for copy-paste), Perlin-noise simulated cloudy S2 data, Perlin-noise cloud coverage maps, copy-paste simulated cloudy S2 data. Columns: three different samples.}
  \label{fig:synthetic}
\end{figure}

To compensate for the lack of any large-scale data set for cloud removal, 
previous work simulated artificial data \cite{Enomoto_Sakurada_Wang_Fukui_Matsuoka_Nakamura_Kawaguchi_2017, Grohnfeldt_Schmitt_Zhu_2018, sintarasirikulchai2018multi, tedlek2018cloud, Rafique_Blanton_Jacobs} of synthetic cloudy optical images. This raises the question of the goodness of the simulated observations, i.e. how good of an approximation such simulations are to any real data. In this experiment we consider the two architectures ours-0 and ours-100 from Table \ref{tab:results} and re-train them on synthetic data to subsequently evaluate the re-trained models on the real test data and assess if performance generalizes to real-world scenarios. Two approaches of generating synthetic data are evaluated: (1) \textit{Perlin:} We generate cloudy imagery via Perlin noise \cite{Perlin} and alpha-blending as in the preceding studies of \cite{Enomoto_Sakurada_Wang_Fukui_Matsuoka_Nakamura_Kawaguchi_2017, Grohnfeldt_Schmitt_Zhu_2018, sintarasirikulchai2018multi}. This approach has the limitation of adding Perlin noise to all of the multi-spectral bandwiths evenly, due to lack of a better physical model of multi-spectral cloud noise. Since cloud detectors trained on real observations are expected to fail in such a case,  we subsitute the cloud map of section \ref{sub:cloudmap} by the synthesized alpha-weighted Perlin noise. 
(2) \textit{Copy:} We generate cloudy imagery by taking the ground-truth cloud-free optical observations and combine them via alpha-blending with clouded observations as in the approach of \cite{Rafique_Blanton_Jacobs}. Different from \cite{Rafique_Blanton_Jacobs} we benefit from our curated data set and alpha-blend paired cloudy-cloudless observations, whereas the prior study mixed two unrelated images. Moreover, we alpha-blend weighted by the cloud map of section \ref{sub:cloudmap}, whereas the original study alpha-blended via sampled Perlin-noise. We believe these modifications better preserve the spectral properties of real observations and keep cloud distribution statistics closer to that of real data as shown in Fig. \ref{fig:synthetic} and \ref{fig:cloud_mask}.
Furthermore, this allows for synthesizing coverage ranging from semi-transparent to fully occluded clouds, which would be less straightforward on unpaired observations. Exemplary observations generated by both simulation approaches as well as empirical observations are presented in Fig. \ref{fig:synthetic}.

The outcomes of this experiment are presented in Table \ref{tab:synthetic}. For all data simulation approaches, training a network on generated data and subsequently evaluating it on synthetic test data is overestimating the performances on the corresponding real test data. This observation holds for both models evaluated in the experiment.The models display a drop in performance when moving from synthetic to real testing data. The drop being considerably smaller in the case of copy-paste data than for Perlin noise data may be due to the copy-pasted data closer resembling the real data and its underlying statistics of cloud coverage and spectral distributions. In this context it is instructive to investigate spectral distortions by means of SAM, which indicates that models trained and tested on synthetic data are considerably poorer to predict spectral distributions on Perlin-simulated data as compared to the copy-pasted observations, which is arguable more alike to real data in terms of its spectral properties. The findings in this experiment underline the need for synthetic data to closely capture the properties of real data, yet even when real and synthetic observations may be hardly distinguishable by eye (as the examples shown in Fig. \ref{fig:synthetic}) there persist important discrepancies unaccounted for that hinder models trained on synthetic sampled to perform equally on real data.

\section{Discussion}
\label{sec:discussion}

The contribution of our work is in providing a large-scale and global data set for cloud removal and developing a new model for recovering cloud-covered information to highlight the data set's benefits. With over $55$ \% of the earth's land surface covered by clouds \cite{King_Platnick_Menzel_Ackerman_Hubanks_2013} the ability to penetrate cloud coverage is of great interest to the remote community in order to obtain a continuous and seamless monitoring of our planet. While the focus in this work is on providing the first globally sampled multi-modal data set for general-purpose cloud removal, future research should  also address the benefits of cloud removal approaches for particular applications common in remote sensing. 
An example application is in semantic segmentation, which necessitates clear-view observations for accurate land-cover classification. Another, in the context of having consecutive observations over time, would be change or anomaly detection where cloud removal methods may be beneficial particularly for the purpose of early-stage detection, which could otherwise be delayed in the presence of clouds.
A limitation of our proposed cloud removal model is its restriction to work on a subset of the optical observation's spectral bands. While this constraint is required due to the choice of the network architecture as necessitated by our experiments conducted, we are certain that it will be beneficial for future research to consider the full spectral information. To allow for this, our curated global data set is released with all available information for both modalities, including the full spectrum of bands for the optical observations.

\section{Conclusion}
\label{sec:conclusion}

We demonstrated the de-clouding of optical imagery by fusing multi-sensory data, proposed a novel model and released the to our knowledge first global data set combining over a hundred-thousand paired cloudy, cloud-free as well as co-registered SAR sample triplets. Statistical analysis of our data set shows a relatively uniform distribution of cloud coverage, with clear images occurring just as probable as wide and densely occluded ones---indicating the need for flexible cloud removal approaches to potentially handle either case. Our proposed network explicitly models cloud coverage and thus learns to retain cloud-free information while as well being able to recover information of areas covered by wide or dense clouds. 
We evaluated our model on a globally sampled test set and measure the goodness of predictions with recently proposed metrics that capture both prediction quality and coverage of the target distribution. Moreover, we showed that our model benefits from supervised learning on paired training data as provided by our large-scale data set. 
Finally, we evaluated the goodness of synthetically generated data of cloudy-cloudless image pairs and show that great performance on synthetic data may not necessarily translate to equal performance on real data. Importantly, when testing on real data then networks trained on real observations consistently outperform models trained on synthetic observations, indicating the existence of properties of the real observations not modeled sufficiently well by the simulated data. This underlines the needs for a set of real observations numerous enough to train large models on, as provided by the data set released in this work. In further studies, we will address the fusion of multi-temporal and multi-sensory data, combining and comparing across both currently segregated approaches. To support future research and make contributions comparable, we share our global data set of paired cloudy, cloud-free as well as co-registered SAR imagery and provide our test data split for benchmarking purposes.

\appendices

\section{Cloud detection} \label{appendix:a}

\begin{figure}[h!tb]
  \centering
  \begin{subfigure}[b]{0.24\linewidth}
    \includegraphics[width=\linewidth]{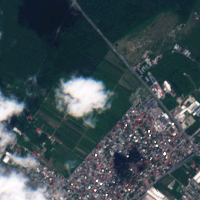}
  \end{subfigure}
  \begin{subfigure}[b]{0.24\linewidth}
    \includegraphics[width=\linewidth]{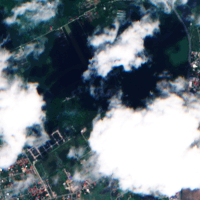}
  \end{subfigure}
  \begin{subfigure}[b]{0.24\linewidth}
    \includegraphics[width=\linewidth]{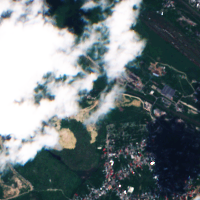}
  \end{subfigure}
  \begin{subfigure}[b]{0.24\linewidth}
    \includegraphics[width=\linewidth]{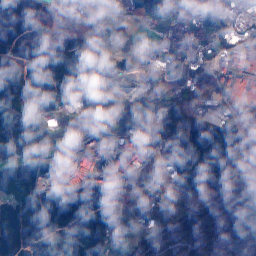}
  \end{subfigure}
    \begin{subfigure}[b]{0.24\linewidth}
    \includegraphics[width=\linewidth]{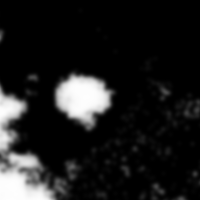}
  \end{subfigure}
  \begin{subfigure}[b]{0.24\linewidth}
    \includegraphics[width=\linewidth]{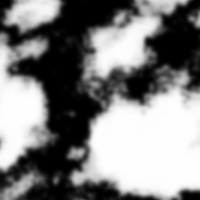}
  \end{subfigure}
    \begin{subfigure}[b]{0.24\linewidth}
    \includegraphics[width=\linewidth]{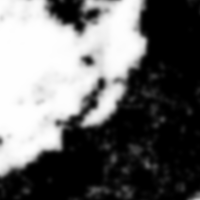}
  \end{subfigure}
  \begin{subfigure}[b]{0.24\linewidth}
    \includegraphics[width=\linewidth]{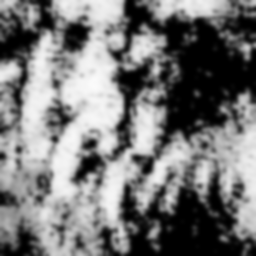}
  \end{subfigure}
  \vskip\baselineskip
  \vskip\baselineskip
  \caption{Exemplary cloudy optical observations and cloud maps. Rows: Cloudy S2 data, cloud probability masks. Columns: Four different samples.}
  \label{fig:cloud_mask}
\end{figure}

We present exemplary cloudy optical observations and cloud maps in Fig. \ref{fig:cloud_mask}. The cloud masks are as predicted by our cloud detection pipeline detailed in section \ref{sub:cloudmap}. The illustrated examples show that our proposed method can reliably detect clouds and provide continuous-valued cloud masks.

\section{Improved precision and recall} \label{appendix:b}
  
  We provide a definition of improved precision and recall in line with the definitions in \cite{Kynkaanniemi_Karras_Laine_Lehtinen_Aila_2019}. For further details the interested reader is referred to the original publication. \linebreak
  
  \textbf{Definition:} \textit{Improved precision and recall \cite{Kynkaanniemi_Karras_Laine_Lehtinen_Aila_2019}}. \newline Let $X_r \sim P_r$ and $X_g \sim  P_g$ denote paired samples drawn from the real and generated distributions of cloud-free images, where $P_g$ is the distribution learned by the generator network whose quality is to be assessed. Each sample is mapped via an auxiliary pre-trained network $M$\footnote{here: VGG16 \cite{Simonyan_Zisserman_2014}, with features extracted at the second fully-connected layer, as argued for in \cite{brock2018large}. Metric evaluation on alternative pre-trained networks has shown to provide virtually identical results \cite{Kynkaanniemi_Karras_Laine_Lehtinen_Aila_2019}.} in a high-dimensional feature space to obtain latent representations $\phi_r =  M(X_r)$ and $\phi_g =  M(X_g)$ such that the two sets of samples are mapped into two feature sets $\Phi_r$ and $\Phi_g$. A distribution $P \in \{P_r, P_g\}$ is approximated by computing pairwise distances between feature embeddings of the observed samples $\Phi \in \{\Phi_r, \Phi_g\}$ and, centered at each feature $\phi \in \Phi$, forming a hypersphere with a radius corresponding to the distance to its k-th nearest neighbor embedding $N_k(\phi)$. Hence, whether an embedded sample $\phi$ falls on manifold $\Phi$ or not is given via
  $$f(\phi,\Phi) = \begin{cases} 1 & \text{if} \: \exists \phi' \in \Phi: ||\phi-\phi'|| \leq ||\phi' - N_k(\phi')||_2 \\
                     0 &  \text{else}
                    \end{cases}$$ 
  The fraction of samples that fall on the the paired distribution's manifold are then defined in \cite{Kynkaanniemi_Karras_Laine_Lehtinen_Aila_2019} as
  $$precision(\Phi_r, \Phi_g) = \frac{1}{|\Phi_g|} \sum_{\phi_g \in \Phi_g} f(\phi_g, \Phi_r)$$
  $$recall(\Phi_r, \Phi_g) = \frac{1}{|\Phi_r|} \sum_{\phi_r \in \Phi_r} f(\phi_r, \Phi_g)$$
  We set parameters $|\Phi| = 7893$ corresponding to the size of the test split of SEN12MS-CR and $k=10$, because every sample has up to 50\% overlap with its neighboring samples. This setting removes the paired target itself plus its eight overlapping samples when computing $N_k(\phi)$.

\section{Cloud coverage statistics on test split} \label{appendix:c}

\begin{figure}[h!tb]
    \includegraphics[width=\linewidth]{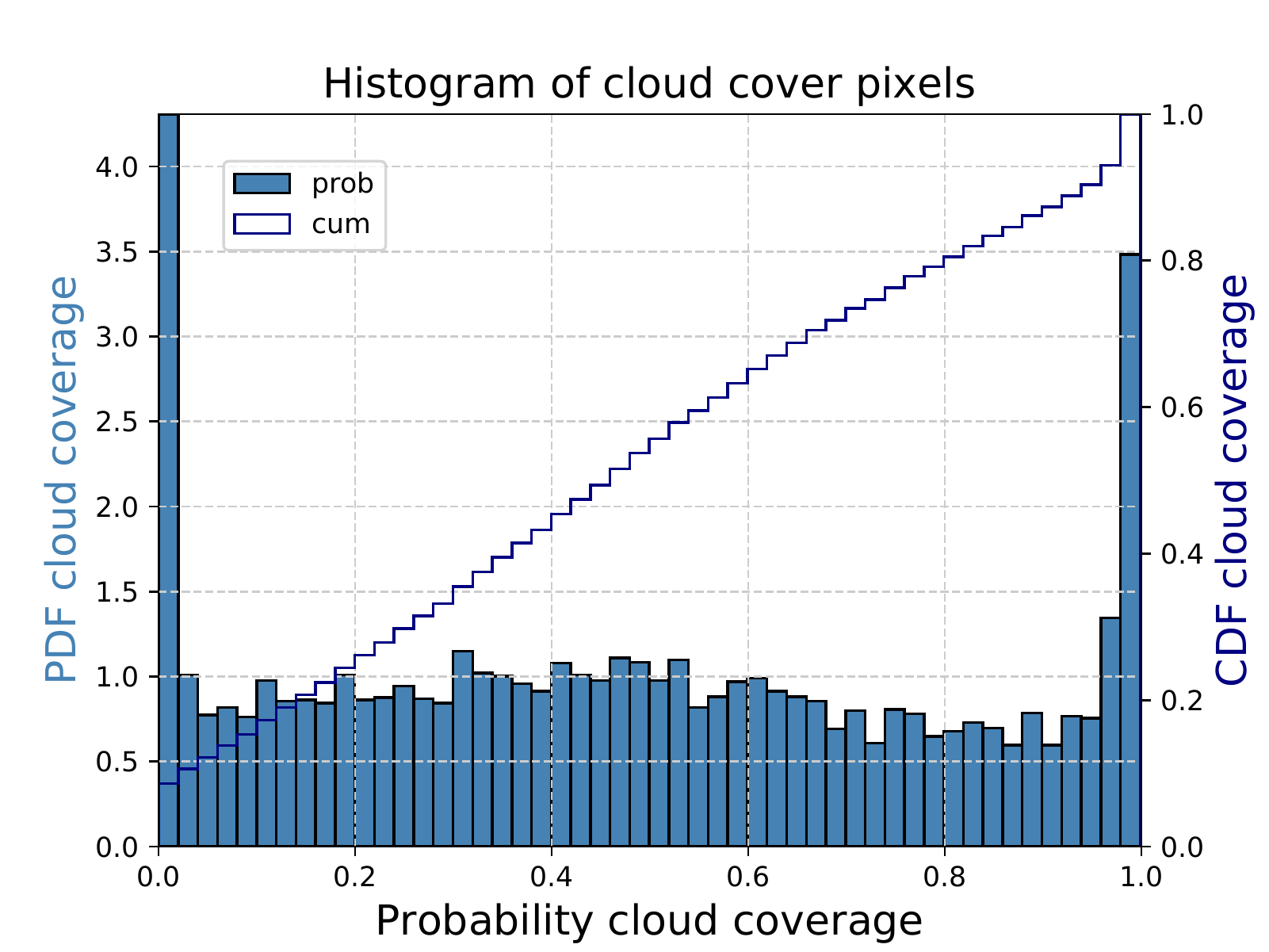}
    \caption{Statistics of cloud coverage of test split of SEN12MS-CR. As for the statistics on the complete data set, an average of circa $50 \%$ of occlusion is observed.}
    \label{fig:cloud_coverage_test}
\end{figure}

In addition to the cloud coverage statistics on the entire data set, as reported in section \ref{data}, Fig. \ref{fig:cloud_coverage_test} provides the empirically observed distribution of cloud coverage on the data set's test split. Even though the histogram of the test split is less smooth than that of the complete data set due to the test split being much smaller, both distributions are considerably alike.

\section{Exemplary problematic cases} \label{appendix:d}

\begin{figure}[h!tb]
  \centering
  \begin{subfigure}[b]{0.3\linewidth}
    \includegraphics[width=\linewidth]{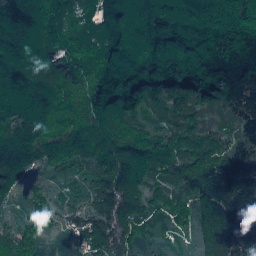}
  \end{subfigure}
  \begin{subfigure}[b]{0.3\linewidth}
    \includegraphics[width=\linewidth]{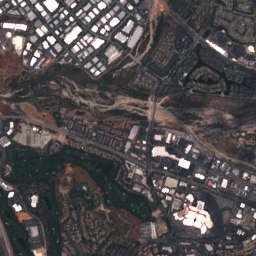}
  \end{subfigure}
    \begin{subfigure}[b]{0.3\linewidth}
    \includegraphics[width=\linewidth]{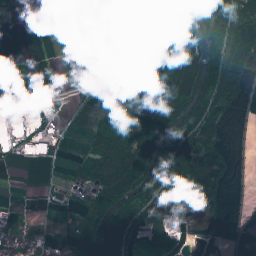}
  \end{subfigure}
  \label{fig:coffee}
  \begin{subfigure}[b]{0.3\linewidth}
    \includegraphics[width=\linewidth]{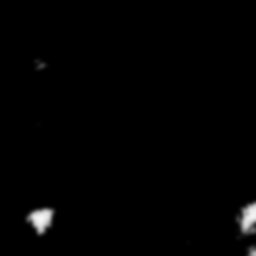}
  \end{subfigure}
  \begin{subfigure}[b]{0.3\linewidth}
    \includegraphics[width=\linewidth]{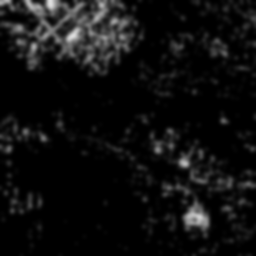}
  \end{subfigure}
    \begin{subfigure}[b]{0.3\linewidth}
    \includegraphics[width=\linewidth]{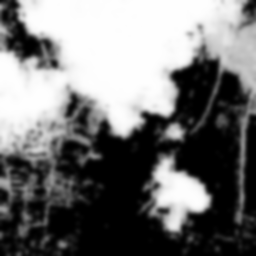}
  \end{subfigure}
  \begin{subfigure}[b]{0.3\linewidth}
    \includegraphics[width=\linewidth]{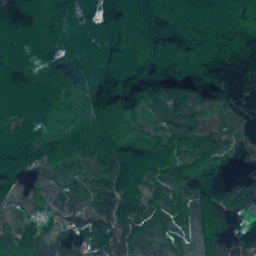}
  \end{subfigure}
  \begin{subfigure}[b]{0.3\linewidth}
    \includegraphics[width=\linewidth]{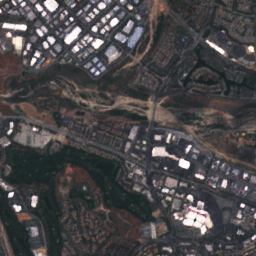}
  \end{subfigure}
    \begin{subfigure}[b]{0.3\linewidth}
    \includegraphics[width=\linewidth]{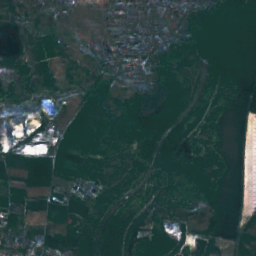}
  \end{subfigure}
  \begin{subfigure}[b]{0.3\linewidth}
    \includegraphics[width=\linewidth]{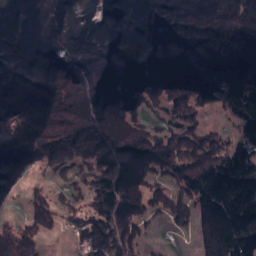}
  \end{subfigure}
  \begin{subfigure}[b]{0.3\linewidth}
    \includegraphics[width=\linewidth]{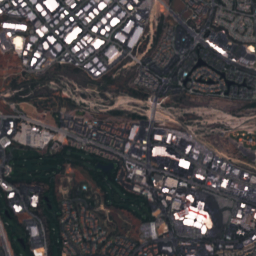}
  \end{subfigure}
    \begin{subfigure}[b]{0.3\linewidth}
    \includegraphics[width=\linewidth]{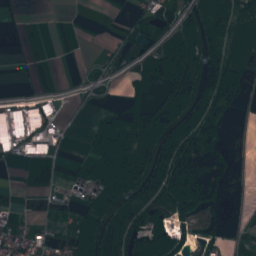}
  \end{subfigure}
  \caption{Exemplary cases posing challenges to our cloud-removal approach. Rows: S2 data (in RGB), predicted cloud map $m$, predicted $\hat{S2}$ data, cloud-free (target) S2 data. Columns: three different samples. Reconstructing optical information obscured by clouds is a hard problem. Among the challenges faced by cloud removal approaches may be: (1) over-time changes in landcover, atmosphere, day time acquisition or seasonality. (2) precise detection of clouds with few misses and false alarms. (3) correct reconstruction of information fully covered by large and dense clouds. 
  }
  \label{fig:inspect}
\end{figure}

For the sake of completeness we discuss cases that we consider challenging for cloud removal approaches, specifically our method, and present exemplary data and predictions of such cases in Fig. \ref{fig:inspect}. We consider the following challenges: (1) Changes in landcover, atmosphere, day time acquisition or seasonality that may occur between (visible parts of) the cloudy reference image and the cloud-free target optical image. While our data set is curated to minimize such cases by selecting observations that are close in time, strict ground truth correspondence is challenging to establish and may only be guaranteed by simulating synthetic data as in experiment \ref{subsub:synthetic}. (2) Precise detection of clouds and accurate cloud masks that minimizes false alarms and misses. With respect to our cloud detection algorithm there exist cloud masks where, even for completely cloud-free images, pixels are assigned a non-zero (albeit rather low) probability of being cloudy. (3) Correct reconstruction of cloud-covered information. In particular for the case of complete coverage by large and dense clouds, this is a very challenging problem. We observed cases where the information reconstructed by our model did not match the target image's, for instance urban-like landcover was predicted in place of agricultural areas.

\section*{Acknowledgment}

The authors would like to thank ESA and the Copernicus program for making the Sentinel observations accessed for this submission publicly available. We would finally like to thank Lloyd Hughes for having shared his artifact detection pre-processing code with us.

\ifCLASSOPTIONcaptionsoff
  \newpage
\fi



\bibliographystyle{IEEEtran}
%\bibliography{../bib/IEEEabrv,../bib/IEEEexample}
\bibliography{references}

%\begin{thebibliography}{1}


%\end{thebibliography}


% commented out until completion / camera-ready version
\begin{IEEEbiography}[{\includegraphics[width=1in,height=1.25in,clip,keepaspectratio]{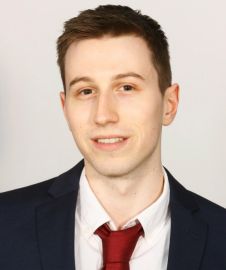}}]{Patrick Ebel} (patrick.ebel@tum.de) received the B.Sc. degree in cognitive science from University of Osnabrück, Germany, in 2015, the M.Sc. degree in Cognitive Neuroscience and the M.Sc degree in Artificial Intelligence from Radboud University Nijmegen, The Netherlands, in 2018. He is currently pursuing the Ph.D. degree with the SiPEO lab, Department of Aerospace and Geodesy of Technical University of Munich, Germany. His research interests include deep learning and its applications in computer vision and to remote sensing data. \end{IEEEbiography}

\begin{IEEEbiography}[{\includegraphics[width=1in,height=1.25in,clip,keepaspectratio]{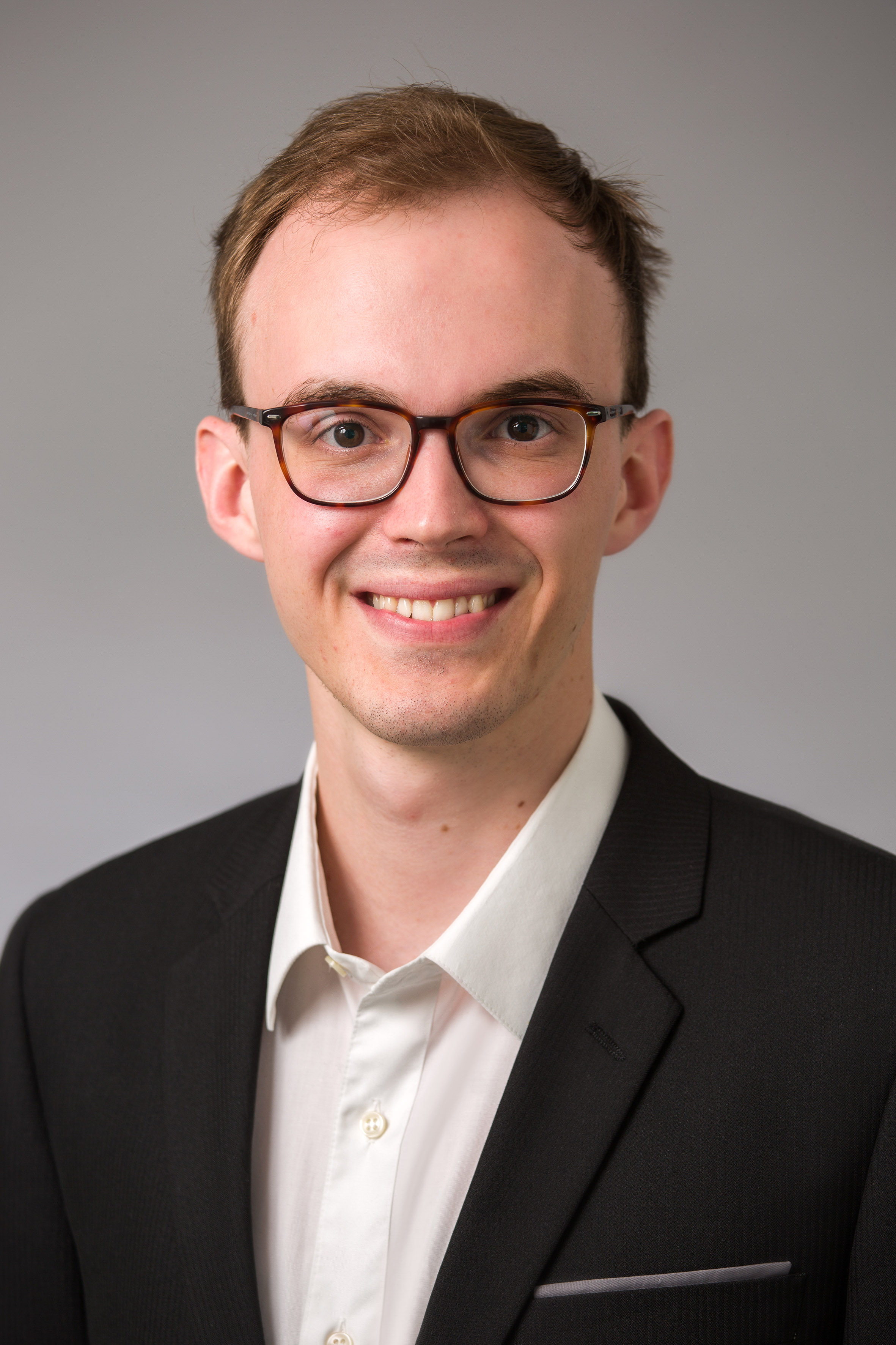}}]{Andrea Meraner} Andrea Meraner received the B.Sc. degree in physics and the M.Sc. degree with high distinction in Earth Oriented Space Science and Technology (ESPACE) at the Technical University of Munich (TUM), Munich, Germany, in 2016 and 2019 respectively. In 2017 he was research assistant in the atmospheric modelling group of the German Geodetic Research Institute (DGFI-TUM) and in 2018 at the German Aerospace Center (DLR) - German Remote Sensing Data Center, Weßling, Germany, in the International Ground Segment department. After spending one semester at the Indian Institute of Technology Mandi, Himachal Pradesh, India, in 2019 he was research assistant at the Signal Processing in Earth Observation (SiPEO) group of TUM and Remote Sensing Technology Institute of DLR, working on deep learning-based cloud removal algorithms for optical satellite imagery. Since October 2019, he is a Junior Remote Sensing Scientist for optical imagery at EUMETSAT, the European Organization for the Exploitation of Meteorological Satellites, in Darmstadt, Germany, developing algorithms to process and analyse data from current and future geostationary satellite missions.\end{IEEEbiography}

\begin{IEEEbiography}[{\includegraphics[width=1in,height=1.25in,clip,keepaspectratio]{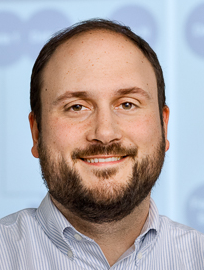}}]{Michael Schmitt}(michael.schmitt@hm.edu) received his Dipl.-Ing. (Univ.) degree in geodesy and geoinformation, his Dr.-Ing. degree in remote sensing, and his habilitation in data fusion from the Technical University of Munich (TUM), Germany, in 2009, 2014, and 2018, respectively.
Since 2020, he has been a full professor for applied geodesy and remote sensing at the Munich University of Applied Sciences, Department of Geoinformatics. From 2015 to 2020, he was a senior researcher and deputy head at the Professorship for Signal Processing in Earth Observation at TUM; in 2019 he was additionally appointed as Adjunct Teaching Professor at the Department of Aerospace and Geodesy of TUM. In 2016, he was a guest scientist at the University of Massachusetts, Amherst. His research focuses on image analysis and machine learning applied to the extraction of information from multi-modal remote sensing observations. In particular, he is interested in remote sensing data fusion with a focus on SAR and optical data. He is a co-chair of the Working Group ``SAR and Microwave Sensing'' of the International Society for Photogrammetry and Remote Sensing, and also of the Working Group ``Benchmarking'' of the IEEE-GRSS Image Analysis and Data Fusion Technical Committee. He frequently serves as a reviewer for a number of renowned international journals and conferences and has received several Best Reviewer awards. He is a Senior Member of the IEEE and an associate editor of IEEE Geoscience and Remote Sensing Letters.
\end{IEEEbiography}

\begin{IEEEbiography}[{\includegraphics[width=1in,height=1.25in,clip,keepaspectratio]{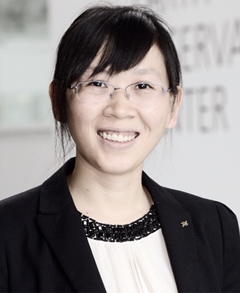}}]{Xiao Xiang Zhu}(S'10--M'12--SM'14) received the Master (M.Sc.) degree, her doctor of engineering (Dr.-Ing.) degree and her “Habilitation” in the field of signal processing from Technical University of Munich (TUM), Munich, Germany, in 2008, 2011 and 2013, respectively.
\par
She is currently the Professor for Signal Processing in Earth Observation (www.sipeo.bgu.tum.de) at Technical University of Munich (TUM) and the Head of the Department ``EO Data Science'' at the Remote Sensing Technology Institute, German Aerospace Center (DLR). Since 2019, Zhu is a co-coordinator of the Munich Data Science Research School (www.mu-ds.de). Since 2019 She also heads the Helmholtz Artificial Intelligence Cooperation Unit (HAICU) -- Research Field ``Aeronautics, Space and Transport". Since May 2020, she is the director of the international future AI lab "AI4EO -- Artificial Intelligence for Earth Observation: Reasoning, Uncertainties, Ethics and Beyond", Munich, Germany. Prof. Zhu was a guest scientist or visiting professor at the Italian National Research Council (CNR-IREA), Naples, Italy, Fudan University, Shanghai, China, the University  of Tokyo, Tokyo, Japan and University of California, Los Angeles, United States in 2009, 2014, 2015 and 2016, respectively. Her main research interests are remote sensing and Earth observation, signal processing, machine learning and data science, with a special application focus on global urban mapping.

Dr. Zhu is a member of young academy (Junge Akademie/Junges Kolleg) at the Berlin-Brandenburg Academy of Sciences and Humanities and the German National  Academy of Sciences Leopoldina and the Bavarian Academy of Sciences and Humanities. She is an associate Editor of IEEE Transactions on Geoscience and Remote Sensing. \end{IEEEbiography}


%\newpage



%\vfill


\end{document}